\documentclass[a4paper,twocolumn,11pt,unpublished]{quantumarticle}
\pdfoutput=1
\usepackage[T1]{fontenc}
\usepackage[utf8]{inputenc}
\usepackage[english]{babel}
\usepackage{amsmath,amssymb,amsthm,mathtools}
\usepackage{graphicx,booktabs,array,longtable,enumitem,needspace}
\usepackage{placeins}
\usepackage[numbers,sort&compress]{natbib}
\usepackage{microtype}
\usepackage{xurl}
\usepackage{hyperref}
\hypersetup{colorlinks=true,linkcolor=blue,citecolor=blue,urlcolor=blue,
  pdftitle={Initialization masks overcome decoding obstructions in compressed QAOA},
  pdfsubject={SEMI-QAOA research manuscript}}
\newtheorem{theorem}{Theorem}[section]
\newtheorem{lemma}[theorem]{Lemma}

\theoremstyle{definition}

\newtheorem{example}{Example}[section]
\theoremstyle{remark}

\numberwithin{equation}{section}
\providecommand{\ket}[1]{\lvert #1\rangle}
\providecommand{\bra}[1]{\langle #1\rvert}
\newcommand{\sgnzero}{\operatorname{sign}_0}
\newcommand{\OPT}{\operatorname{OPT}}
\newcommand{\E}{\mathbb E}
\newcommand{\Var}{\operatorname{Var}}
\newcommand{\Cut}{\operatorname{Cut}}
\newcommand{\calD}{\mathcal D}
\newcommand{\allplus}{\mathbf 1}

\newcommand{\SEMI}{\mathrm{SEMI}}
\newcommand{\encoder}{\texttt{exact\_\allowbreak coloring\_\allowbreak chunked\_\allowbreak rot\_\allowbreak v1}}
\setlist{itemsep=2pt,topsep=4pt,parsep=0pt}
\makeatletter
\patchcmd{\@maketitle}{\begin{minipage}{\textwidth}}{\begin{minipage}{\textwidth}\raggedright}{}{}
\makeatother
\begin{document}
\normalfont
\raggedbottom
\title{Initialization masks overcome decoding obstructions in compressed QAOA}
\author{Emilio Semre}
\affiliation{Faculty of Computer Science, Technion--Israel Institute of Technology, Haifa, Israel}
\author{Steven Frankel}
\affiliation{Faculty of Mechanical Engineering, Technion--Israel Institute of Technology, Haifa, Israel}
\maketitle
\begin{abstract}
In compressed quantum optimization, a variational state is converted into a
classical assignment by decoding noncommuting observables. We show that this
step creates structural restrictions that changing the initialization can
remove even when increasing circuit depth cannot. For an explicit infinite
family of connected cubic graphs on $n=3r$ vertices, with a fixed Pauli
encoding and deterministic sign decoder, every all-plus QAOA circuit has cut
at most $2r$, at every finite depth and every real angle sequence. A uniform
signed product-state mask at fixed depth-one angles has expected cut
$13r/4$, an additive advantage $5n/12$; an infinite subfamily admits an
optimal mask. A second construction retains a linear depth-one separation
under weak weighted couplings, a bounded cost-angle domain and a specified
production-encoder ordering.
Noiseless simulations then test classical mask selection before energy
optimization. On 1,699 random cubic MaxCut instances, one selected mask
achieves mean ratio $0.9100$, versus $0.7067$ for the all-plus $X$ baseline
and $0.9125$ for the best of $r^2$ separately optimized random masks.
On complete cohorts of weighted MaxCut, sparse Ising, Max-2-XOR and
Max-2-SAT, the selected mask also improves on the baseline at every size,
but it trails the pool at small sizes and leads it significantly only at
the larger MaxCut and Ising sizes. Finite-pool order statistics separate the
effect of selection from that of drawing a random mask. The results
distinguish decoded expressivity from optimized energy and empirical
performance; they do not establish a quantum computational advantage.
\end{abstract}
\section{Introduction}
\label{sec:introduction}
Quantum random access optimization (QRAO) places several binary variables
on different Pauli axes of one qubit, replacing a classical objective by a
noncommuting quantum relaxation~\cite{fuller2024}. Combining this encoding
with the quantum approximate optimization algorithm (QAOA)~\cite{farhi2014}
reduces the register size, but separates two optimization tasks: preparing a
state with favorable relaxed energy and decoding a good classical solution.
An ansatz can explore many quantum states while its deterministic decoder
remains confined to a much smaller set of assignments.

This paper identifies an obstruction of that kind and a way to escape it.
If the reference state, cost Hamiltonian and mixer share a physical
permutation symmetry, every circuit depth preserves that symmetry. A
deterministic equivariant decoder then returns an assignment fixed by the
same permutation. Changing the signs of a product-$X$ reference state,
together with its matched mixer, can remove this restriction without changing
the encoding or the number of variational angles. The resulting discrete
initialization choice is called a \emph{mask}.

Matched initializations and mixers are established elements of variational
algorithm design~\cite{hadfield2019,egger2021,he2023alignment}. For compressed
optimization, He \emph{et al.} compared matched $X$, $Y$ and $Z$ choices
\cite{he2025}. Rand-SEMI-QAOA evaluated finite random pools of independently
optimized signed product-$X$ masks with matched mixers~\cite{semre2026randsemi}.
We use that family rather than introduce a new ansatz. Our contribution concerns what its masks make \emph{decodable},
and how much of their empirical benefit classical selection can recover
with one subsequent angle optimization.

\paragraph{Contributions and assumptions.}
The principal result is an extensive separation on explicitly encoded
connected cubic graphs: every all-plus circuit, at arbitrary finite depth
and arbitrary real angles, decodes a cut at most $2r$, whereas a uniform
mask at one fixed depth-one angle has mean cut $13r/4$. An infinite
subfamily admits a mask cutting every edge (Theorem~\ref{thm:cubic}).
A second theorem proves a linear depth-one separation for weakly coupled
weighted modules under a specified ordering of the production encoder
(Theorem~\ref{thm:weighted}). Both use expectation-sign decoding, including
the fixed convention $\sgnzero(0)=+1$. The first theorem is specific to its
displayed packing; the second has an explicit bounded cost-angle domain.

The numerical study holds the encoding, decoder and angle-optimization
protocol fixed while comparing classical selectors with the all-plus $X$
baseline and an $r^2$-mask random pool. On 1,699 MaxCut instances, MARS-SEMI
improves the mean decoded ratio by $0.2033$ over all-plus, with reported
95\% bootstrap interval $[0.1993,0.2075]$. Complete cohorts of weighted
MaxCut, sparse Ising and Max-2-XOR, with weighted Max-2-SAT as supplementary
evidence, test how this result depends on the objective. Against the pool,
the selected mask trails at small sizes in every family and leads at the
larger MaxCut and Ising sizes; for the other objectives the difference
becomes nonsignificant without an established reversal. Finite-pool comparisons isolate
selection from merely allowing random masks. We do not identify the
constructed symmetry obstruction as the cause of the random-graph
improvements.

\paragraph{Relation to Rand-SEMI-QAOA.}
Reference~\cite{semre2026randsemi} defined the product-mask pools, derived
their gauge description and the finite-population best-of-$k$ accounting,
and evaluated the pools on 2,400 random cubic MaxCut graphs with
$n\in\{18,20,22\}$. Its theory does not order the
decoded maxima of different masks. The present work is distinct in three
respects: the exact decoded separations between all-plus and masked circuits
(Example~\ref{ex:k33} and Theorems~\ref{thm:cubic} and~\ref{thm:weighted});
classical selectors that choose one mask before a single angle optimization;
and their comparison with both the all-plus baseline and the $K=r^2$ random
pool on five objectives with up to $n=50$ variables, using the retained
MaxCut benchmark and complete extension cohorts.
The random pool is rerun here under the common protocol. The 300 MaxCut
graphs with $n=18,20,22$ belong to the earlier benchmark, and the weighted
MaxCut, Ising and independent-set instances at $n=20$ reuse its topologies
with different objectives; these cells are not independent replications.

The all-plus $X$ comparator is the specific baseline studied here, not the
best of all QRAO initializations or rounding rules. Our claims concern
decoded expressivity and finite-size numerical performance, not superiority
to competitive classical optimization.

\section{Encoding, masks and the decoded objective}
\label{sec:framework}
\begin{figure*}[t]
 \centering
 \includegraphics[width=\textwidth]{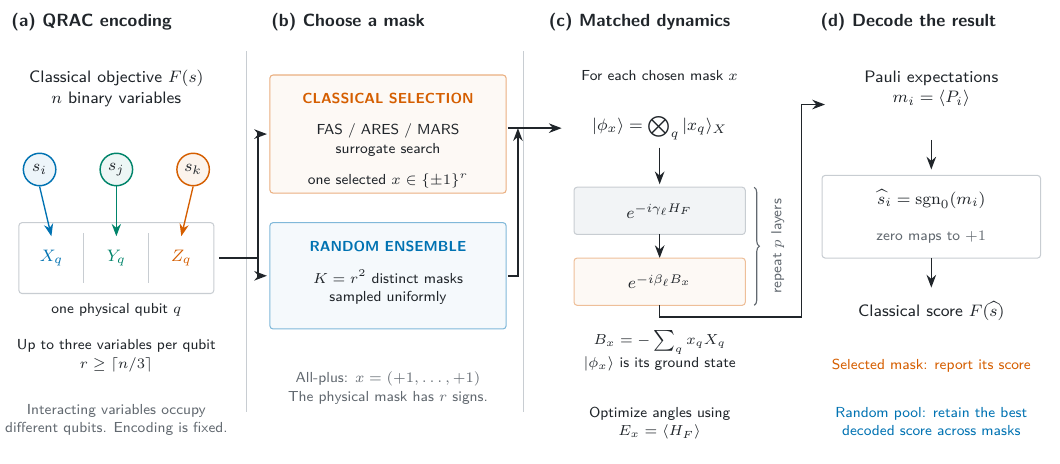}
 \caption{\textbf{SEMI-QAOA: a discrete initialization choice inside a
 compressed variational pipeline.} An admissible $(3,1)$-QRAC encoding
 assigns up to three nonadjacent logical variables to different Pauli axes
 of each physical qubit and fixes the relaxed Hamiltonian $H_F$.
 A mask $x\in\{\pm1\}^r$ specifies both the product-$X$ reference
 $\ket{\phi_x}$ and its matched mixer $B_x=-\sum_qx_qX_q$.
 Cost and mixer layers are alternated, with angles optimized for relaxed
 energy. Pauli expectations are then decoded by the common sign rule,
 including $\sgnzero(0)=+1$, and scored in the original classical objective.
 The all-plus baseline optimizes one fixed mask; a classical selector
 chooses one mask before angle optimization; Rand-SEMI optimizes
 $K=r^2$ distinct uniformly sampled masks separately and retains the best
 decoded score. Classical selection has its own computational cost, and
 one optimized mask entails many circuit evaluations. The numerical study
 uses $p=1$ and noiseless expectations.}
 \label{fig:architecture}
\end{figure*}

Figure~\ref{fig:architecture} summarizes the common circuit and the
distinct choices of initialization budget used below.
For binary signs $s_i\in\{\pm1\}$, consider
\begin{equation}
 F(s)=c+\sum_i h_i s_i+\sum_{ij\in E}J_{ij}s_i s_j.
 \label{eq:objective}
\end{equation}
An admissible unsigned $(3,1)$-QRAC encoding assigns variable $i$ to
$P_i\in\{X_{q(i)},Y_{q(i)},Z_{q(i)}\}$ on an $r$-qubit register.
Each axis contains at most one variable, and every interaction joins
different qubits. The relaxed Hamiltonian is
\begin{equation}
 H_F=cI+\sqrt3\sum_i h_iP_i+3\sum_{ij\in E}J_{ij}P_iP_j.
 \label{eq:relaxation}
\end{equation}
The factors remain $\sqrt3$ and $3$ on partially occupied qubits.
For a cut with positive edge weights $w_{ij}$,
\begin{equation}
 \begin{aligned}
 C_G(s)&=\sum_{ij\in E}w_{ij}\frac{1-s_i s_j}{2},\\
 H_C&=\frac12\sum_{ij\in E}w_{ij}(I-3P_iP_j).
 \end{aligned}
 \label{eq:cut}
\end{equation}
The scalar term affects only a global phase.

A physical mask $x\in\{\pm1\}^r$ defines the matched pair
\begin{equation}
 \ket{\phi_x}=\bigotimes_q\ket{x_q}_X,
 \qquad B_x=-\sum_qx_qX_q,
 \label{eq:mask}
\end{equation}
where $\ket{\phi_x}$ is the unique ground state of $B_x$. With the cost
layer preceding its mixer, the depth-$p$ state is
\begin{equation}
 \ket{\psi_{x,p}}=
 \prod_{\ell=p}^{1}\left(e^{-i\beta_\ell B_x}
 e^{-i\gamma_\ell H_F}\right)\ket{\phi_x}.
 \label{eq:circuit}
\end{equation}
The product is ordered with layer one rightmost. The all-plus mask
$x=\allplus$ is called QRAC-QAOA in the numerical comparisons; SEMI-QAOA
allows the mask to vary. The binary label zero in the software denotes
this all-plus mask. The mixer sign in Eq.~\eqref{eq:mask} is fixed
throughout; changing it requires reversing every mixer angle.

For $m_i=\langle\psi_{x,p}|P_i|\psi_{x,p}\rangle$, decode
\begin{equation}
 \widehat s_i=\sgnzero(m_i),\qquad
 \sgnzero(u)=\begin{cases}+1,&u\ge0,\\-1,&u<0.\end{cases}
 \label{eq:decoder}
\end{equation}
Write $F_{x,p}=F(\widehat s)$ and $C_{x,p}$ for a cut. The same rule
applies to all variables, including exact-zero marginals.

Three objects must be distinguished. \emph{Decoded expressivity} maximizes
$F_{x,p}$ over the allowed parameters. For the depth-one theorems, let
\begin{equation}
 \mathcal D=[0,\pi]\times[-\pi,63\pi/65],\qquad
 \theta=(\beta,\gamma),
 \label{eq:domain}
\end{equation}
and define
\begin{equation}
 \begin{aligned}
 F^*_{+,1}&=\max_{\theta\in\mathcal D}F_{\allplus,1}(\theta),\\
 F^*_{\SEMI,1}&=\max_{\substack{x\in\{\pm1\}^r\\\theta\in\mathcal D}}
 F_{x,1}(\theta).
 \end{aligned}
 \label{eq:expressivity}
\end{equation}
These maxima exist because the attainable classical values form a finite
nonempty set. In contrast, \emph{energy optimization} chooses angles using
$E_x(\theta)=\langle\psi_x(\theta)|H_F|\psi_x(\theta)\rangle$ and reports
$F_{x,1}(\theta_x^{\rm opt})$. Even a global energy maximum need not
maximize the decoded objective. Finally, a \emph{pool} reports
$\max_{x\in\mathcal P}F_{x,1}(\theta_x^{\rm opt})$, after optimizing
each candidate separately. A fixed-angle witness is therefore not a
performance guarantee for the experimental optimizer.

Encodings are frozen before comparing masks. The production encoder,
\encoder{}, uses a minimum proper coloring with deterministic tie ordering,
sorts each color class, chunks it into triples, and assigns position $j$
in qubit $q$ the axis $\texttt{XYZ}[(j+q)\bmod3]$ (zero-based indices).
The cubic theorem instead specifies its packing directly; the weighted
theorem specifies the labels and ordering that reproduce its packing under
this encoder. Re-encoding is outside either comparison.

\section{Structural separations}
\label{sec:separations}
\subsection{Why a mask can matter more than depth}
Suppose a physical unitary permutes the logical observables and leaves
the reference, Hamiltonian and mixer invariant. Every circuit layer then
preserves the corresponding state symmetry. Expectation-sign decoding is
equivariant under this permutation, so its output is a labeled assignment
fixed by that permutation. This restriction holds for all angles and depths, including
at zero marginals. A mask can change the symmetry of the reference and
mixer while leaving the encoded problem fixed.

\begin{example}[A fully packed $K_{3,3}$]
\label{ex:k33}
Use the fully packed $K_{3,3}$ encoding illustrated in Ref.~\cite{he2025}:
place each bipartition on one qubit, using all three unsigned Pauli axes.
Writing $A=X+Y+Z$ gives $H_C=9I/2-3A_1A_2/2$. The all-plus reference,
cost and mixer are invariant under exchanging the qubits. Their decoded
triples must agree, say $(a,a)$, and the cut is
$[9-(\sum_j a_j)^2]/2\in\{0,4\}$. The mask $(+1,-1)$ at
$(\beta,\gamma)=(\pi/8,-\pi/18)$ instead gives Bloch vectors
$m_1=(1,\sqrt2,\sqrt2)/3$ and $m_2=-m_1$, cutting all nine edges.
At those angles all-plus attains four. Thus the all-depth all-plus
optimum is four and the depth-one SEMI optimum is nine.
\end{example}
Appendix~\ref{app:k33} verifies the witness directly. The conclusion
requires this packing and deterministic, exchange-equivariant decoding
into labeled signs; randomized decoders and unoriented partitions are
outside its scope.

\subsection{An extensive obstruction on connected cubic graphs}
For even $r\ge4$, set $h=r/2$ and use indices modulo $r$. The logical
vertices $X_q,Y_q,Z_q$ occupy their named axes on qubit $q$. Define $G_r$
by the edges
\begin{equation}
 \begin{aligned}
 &\{X_q,X_{q+h}\},&&0\le q<h,\\
 &\{Y_q,Y_{q+1}\},\ \{Z_q,Z_{q+1}\},&&0\le q<r,\\
 &\{X_q,Y_{q+1}\},\ \{X_q,Z_{q+1}\},&&0\le q<r.
 \end{aligned}
 \label{eq:cubic-family}
\end{equation}
This is a simple connected cubic graph with $n=3r$ vertices and $9r/2$
edges. Let $C_r$ be the cycle on physical indices and
$M_h=\{\{q,q+h\}:0\le q<h\}$ its antipodal matching.

\begin{theorem}[An all-depth decoded separation]
\label{thm:cubic}
For Eq.~\eqref{eq:cubic-family} and its displayed encoding,
\begin{equation}
 \sup_{\substack{p\ge1\\
 \boldsymbol\beta,\boldsymbol\gamma\in\mathbb R^p}}
 C_{\allplus,p}(\boldsymbol\beta,\boldsymbol\gamma)=2r.
 \label{eq:cubic-cap}
\end{equation}
At
$\theta_*=(\tfrac12\arctan4,-1/1080)\in\mathcal D$, every mask satisfies
\begin{equation}
 C_{x,1}(\theta_*)=f_r(x)
 =2r+2\Cut_{C_r}(x)+\Cut_{M_h}(x).
 \label{eq:cubic-score}
\end{equation}
For a uniform mask $X$,
\begin{equation}
 \E f_r(X)=13r/4,\qquad \Var f_r(X)=9r/8.
 \label{eq:cubic-moments}
\end{equation}
The expected advantage over Eq.~\eqref{eq:cubic-cap} is $5n/12$.
If $h$ is odd, the alternating mask attains
$C^*_{\SEMI,1}=\OPT(G_r)=9r/2$, with advantage $5n/6$.
\end{theorem}

\paragraph{Proof mechanism.}
Cyclically shifting the physical qubits preserves every edge family, the
all-plus mixer and the all-plus state. Thus all-plus decoding is constant
on each axis orbit. Every $XX$, $YY$ and $ZZ$ edge is uncut, leaving only
the $2r$ spokes. This argument is independent of the depth.

For the SEMI witness, the leading small-angle terms alone are insufficient:
one must certify the signs at a nonzero angle. Appendix~\ref{app:local-bounds}
proves a local operator-norm remainder bound for the exact evolution.
At $\theta_*$ it yields, for every mask,
\begin{equation}
 \widehat s_{X_q}=x_q,\qquad
 \widehat s_{Y_q}=\widehat s_{Z_q}=-x_{q-1},
 \label{eq:cubic-decoded}
\end{equation}
with transverse absolute marginals at least $1/(144\sqrt{17})$.
All spokes are cut; the cycles and matching give Eq.~\eqref{eq:cubic-score}.
The constant mask attains the all-plus cap. Expanding the remaining cut
indicators into distinct products of two independent signs proves the mean
and variance. When $h$ is odd, the alternating mask cuts both the cycle
and matching, attaining the full edge count. The complete proof, including
the finite-angle norm estimates, is in Appendix~\ref{app:cubic}.

The theorem compares a fixed-angle mask distribution with the \emph{best}
all-plus output at any depth. Its uniformity in $r$ is substantive: neither
the witness angle nor the sign margin shrinks with graph size. The graph
and encoding are nevertheless constructed. The result is not a typical-case
claim for random cubic graphs, and independent energy optimization can move
a mask away from its certified witness.

\subsection{Robust separation under weak weighted coupling}
The preceding obstruction uses an exact physical permutation symmetry.
The next result asks whether a decoded gap can survive interactions between
otherwise identical modules. It gives a separate depth-one, bounded-angle
answer under a specified production ordering.

For $m\ge3$, module $t$ contains $L_{t,a},R_{t,a}$, $a\in\{X,Y,Z\}$,
and the six unit-weight edges $L_{t,a}R_{t,b}$ with $a\ne b$.
For each axis choose an independent uniform derangement $\pi_a$ of the
modules, and add edges $L_{t,a}R_{\pi_a(t),a}$ with independent weights
$w_{t,a}$ uniform on the open interval $1/400<w_{t,a}<1/200$.
Condition on connectivity. Write
$W_a=\sum_t w_{t,a}$ and $W=\sum_aW_a$. The graph is simple, bipartite
and cubic, with $n=6m$. Appendix~\ref{app:weighted} gives the vertex
labels and ascending coloring order for which the production encoder
places each module side on one fully occupied qubit, with $r=2m$.

\Needspace{12\baselineskip}
\begin{theorem}[A robust weighted separation]
\label{thm:weighted}
Under the graph law and production ordering just specified, every connected
sample satisfies
\begin{equation}
 \begin{aligned}
 C^*_{\SEMI,1}&=\OPT=6m+W,\\
 C^*_{+,1}&\le4m+W-\min_aW_a.
 \end{aligned}
 \label{eq:weighted-gap}
\end{equation}
The gap is greater than $(2+1/400)m$. The all-plus bound holds at depth
one for every real $\beta$ and $|\gamma|\le\pi$. The mask
$x_L=+1,x_R=-1$ attains the optimum at
$(\beta,\gamma)=(\pi/4,-1/486)\in\mathcal D$.
\end{theorem}

\paragraph{Proof mechanism.}
Without weak edges, both qubits of every all-plus module have a common
Bloch vector. A conserved component of the cost evolution ensures that
some coordinate has absolute value at least $1/3$, after any mixer angle.
The coordinate can depend on the angles but is the same for all modules.
A module-local Duhamel bound limits the change in each marginal to
$9|\gamma|/100<1/3$ on the stated interval, uniformly in $m$.
All vertices on the selected axis therefore have the same strict decoded
sign. This leaves at least two strong edges per module and all weak edges
on that axis uncut. For the SEMI mask, the finite-angle expansion instead
recovers the bipartition with transverse margins at least $1/324$.
Appendix~\ref{app:weighted} proves these estimates and the encoder statement.

The result shows robustness to the stated weak couplings, not an arbitrary
perturbation or arbitrary-depth separation. Its weight distribution differs
from the numerical weighted-MaxCut ensemble. The two constructions establish
that initialization can change attainable decoded solutions; the following
study asks how effectively classical selection exploits mask freedom on
separate benchmark distributions.

\section{Classical selection on the benchmark cohorts}
\label{sec:experiments}
\subsection{Protocol and comparison budgets}
We now test whether classical selection can exploit mask freedom before
optimizing circuit angles. All reported runs are noiseless statevector
simulations at $p=1$, with no finite-shot or hardware-noise model. For each
instance the methods share the encoding, relaxed Hamiltonian, deterministic
decoder and prescribed per-mask optimizer. The latter evaluates a
$33\times65$ angle grid, then refines three candidates with L-BFGS-B
\cite{byrd1995}, allowing at most 80 iterations per start. Its optimization
objective is the relaxed energy; final comparisons use decoded scores.

The baseline, QRAC-QAOA, optimizes the all-plus mask and its matched $X$
mixer with this encoding, optimizer and decoder. Rand-SEMI optimizes $K=r^2$
distinct masks drawn uniformly at random and retains the best decoded result.
FAS-SEMI selects one
mask using the deterministic-sign first-order surrogate. ARES-SEMI uses
the implemented Adaptive-RLC2 local-cluster surrogate; MARS-SEMI optimizes
the same surrogate with an ARES warm start. These classical routines use
selector angles $(0.418,-0.127)$, after which the chosen mask receives
the full energy optimization. The small-angle proof bounds do not certify
this operating point or these selectors. Appendix~\ref{app:protocol}
specifies the protocol and the implementation information required for
reproducibility.

The first-order score illustrates why a classical mask predictor is
available. For unweighted MaxCut, let $F_{q,a}(x)$ be the sum of mask signs
on logical $X$ neighbors of the variable occupying axis $a$ of qubit $q$,
with $F_{q,a}=0$ for an unoccupied slot. Set $c=\cos2\beta$ and
$t=\sin2\beta$. The expansion proved in
Appendix~\ref{app:local-bounds} gives the predictors
\begin{equation}
 \begin{aligned}
 \ell_{qX}&=x_q,\\
 \ell_{qY}&=3\gamma(tF_{q,Y}-cx_qF_{q,Z}),\\
 \ell_{qZ}&=3\gamma(tF_{q,Z}+cx_qF_{q,Y}).
 \end{aligned}
 \label{eq:selector-predictor}
\end{equation}
Applying $\sgnzero$ to these local expressions gives a classical
assignment and hence the surrogate score
\begin{equation}
 D_1(x)=\sum_{ij\in E}
 \frac{1-\sgnzero(\ell_i(x))\sgnzero(\ell_j(x))}{2}.
 \label{eq:selector-score}
\end{equation}
Evaluating this score does not require the full $r$-qubit statevector.
Searching it is still a classical optimization task. Moreover, a zero
predictor can receive a nonzero higher-order correction, and a mask
preferred by a surrogate can change its decoded assignment after energy
optimization. The local-cluster selectors seek a more informative score;
their recorded performance, rather than the separation theorems, is the
evidence for that choice.

A selected mask therefore entails many circuit evaluations, not one state
preparation or shot. The random pools contain between 36 and 324 separately
optimized masks per instance (Table~\ref{tab:focused-cohorts}), whereas each
selector optimizes one. This
comparison measures whether classical selection can substitute for repeated
candidate optimizations; it does not equalize total computation. Surrogate
construction, search and warm-start costs remain part of the selected-mask
pipeline.

The retained MaxCut benchmark has 1,699 of 1,700 designed random cubic graphs
with $n=18,20,\ldots,50$. The complete extension cohorts are weighted MaxCut
and weighted Max-2-XOR, each on 900
instances with $n=16,20,\ldots,48$; and sparse Ising with fields on 800
instances with $n=16,20,\ldots,44$. Each size contains 100 instances except
MaxCut $n=50$, whose 99 successful instances are retained after one failed
task; this cell is not a complete sample of its design. A weighted Max-2-SAT
family is reported in Appendix~\ref{app:max2sat}, and a constrained
independent-set family in Appendix~\ref{app:feasibility}. The incomplete
cells, weighted MaxCut $n=52$ (55 of 100 instances), Ising $n=48$ (98) and
$n=52$ (61), and Max-2-XOR $n=52$ (46 of a designed 50), appear in the
figures as open symbols or italic entries but are excluded from every pooled
statistic and conclusion. Their campaigns were stopped by walltime limits or
by decision, so their completion can depend on run time.
The cut and XOR scores divide the decoded value by the certified optimum.
Ising instead uses $(F-F_{\min})/(F_{\max}-F_{\min})$, with both extremes
certified. Several families share underlying graph topologies and do not
constitute independent graph replications.

Means weight instances equally; differences pair methods on the same
instance. Reported 95\% percentile bootstrap intervals use 10,000
instance-level resamples~\cite{efron1979}. Per-size intervals are pointwise,
not simultaneous, and are not adjusted for multiplicity. The random masks
within one instance are not additional independent observations.
Appendix~\ref{app:protocol} gives the generators, normalization, seeds and
statistical conventions.

\begin{figure*}[t]
 \centering
 \includegraphics[width=\textwidth]{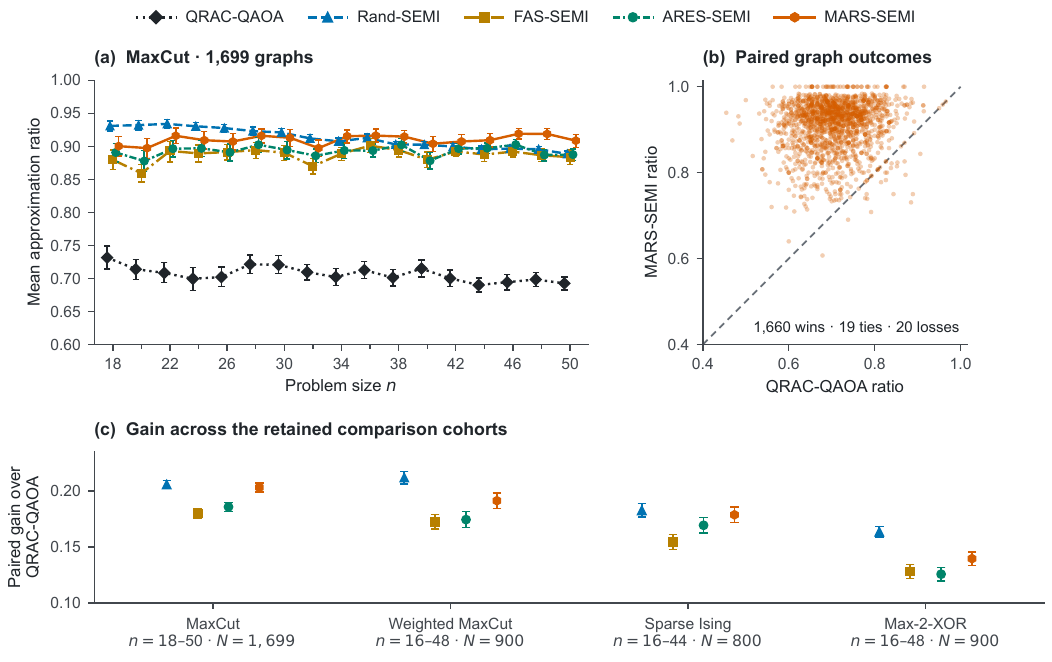}
 \caption{\textbf{Classical mask selection improves decoded quality over
 the all-plus $X$ baseline.} All runs have $p=1$. (a)~Mean MaxCut ratio on
 1,699 random cubic graphs: 100 per size except 99 at $n=50$.
 (b)~MARS-SEMI versus QRAC-QAOA on the same graphs, with 1,660 wins,
 19 ties and 20 losses (tolerance $10^{-12}$); the dashed line is equality.
 (c)~Paired improvements over QRAC-QAOA on the retained benchmark cohorts: MaxCut
 (1,699), weighted MaxCut (900), sparse Ising with fields (800) and weighted Max-2-XOR (900). Ising uses a
 range-normalized objective; the other scores are decoded-value/optimum
 ratios. Bars are pointwise 95\% instance-bootstrap intervals. Rand-SEMI
 optimizes $r^2$ masks; every other method optimizes one.}
 \label{fig:semi-vs-qrac}
\end{figure*}

\subsection{Improvement over all-plus and comparison with a pool}
On MaxCut, the mean ratio increases from $0.7067$ for all-plus to
$0.9100$ for MARS and $0.9125$ for Rand-SEMI (Fig.~\ref{fig:semi-vs-qrac}).
The paired MARS improvement is $0.20331$ with interval
$[0.19926,0.20746]$: 1,660 instances improve, 19 tie and 20 worsen.
FAS and ARES attain means $0.8866$ and $0.8926$, respectively.
The improvement is thus broad but not universal instance by instance.
MARS also improves the mean over all-plus by $0.191$, $0.179$ and
$0.140$ on the complete weighted-MaxCut, Ising and Max-2-XOR cohorts. In
every retained size cell of these four families and of Max-2-SAT, each of
the four mask methods has a paired interval above zero. These are
within-family differences; the absolute scores of different objectives are
not directly comparable.

\begin{table*}[t]
\centering
\small
\caption{Benchmark comparisons at $p=1$, with paired 95\%
instance-bootstrap intervals. The per-size rows classify the MARS-SEMI minus
Rand-SEMI difference in each size cell by its pointwise interval.
MaxCut retains 99 of 100 designed instances at $n=50$;
all other included size cells are complete. Max-2-SAT is reported in
Appendix~\ref{app:max2sat}.}
\label{tab:benchmark-comparisons}
\footnotesize
\setlength{\tabcolsep}{4pt}
\begin{tabular}{@{}llllll@{}}
\toprule
& MaxCut & Weighted MaxCut & Sparse Ising & Max-2-XOR & Max-2-SAT\\
\midrule
Included sizes $n$ & 18--50 & 16--48 & 16--44 & 16--48 & 16--48\\
Instances & 1,699 & 900 & 800 & 900 & 900\\
\addlinespace
MARS $-$ QRAC & $+0.2033$ & $+0.1914$ & $+0.1788$ & $+0.1396$ & $+0.0623$\\
\quad 95\% interval & $[0.1993,0.2075]$ & $[0.1845,0.1984]$ & $[0.1718,0.1859]$ & $[0.1334,0.1457]$ & $[0.0599,0.0647]$\\
MARS $-$ Rand & $-0.0026$ & $-0.0207$ & $-0.0040$ & $-0.0240$ & $-0.0061$\\
\quad 95\% interval & $[-0.0053,+0.0002]$ & $[-0.0250,-0.0164]$ & $[-0.0081,+0.0001]$ & $[-0.0284,-0.0198]$ & $[-0.0073,-0.0049]$\\
\addlinespace
\multicolumn{6}{@{}l}{Per-size MARS $-$ Rand interval, classified by size $n$}\\
\quad below zero & 18--26, 32 & 16--32 & 16, 20 & 16--40 & 16--36\\
\quad contains zero & \begin{tabular}[t]{@{}l@{}}28--30, 34--36,\\40--42\end{tabular} & 36--48 & 24--32 & 44, 48 & 40--48\\
\quad above zero & 38, 44--50 & none & 36--44 & none & none\\
\bottomrule
\end{tabular}
\end{table*}

\begin{figure*}[t]
 \centering
 \includegraphics[width=\textwidth]{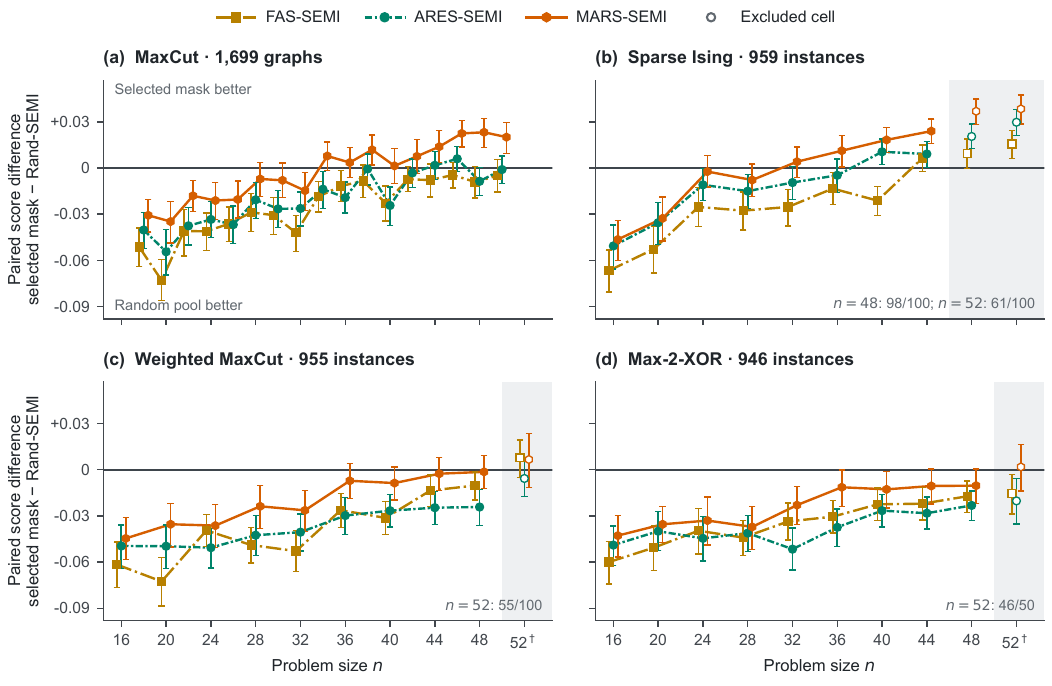}
 \caption{\textbf{One selected mask versus $r^2$ optimized random masks.}
 Paired per-size differences of FAS-, ARES- and MARS-SEMI minus Rand-SEMI;
 positive differences favor the selected mask. (a)~MaxCut, 100 instances
 per size except 99 at $n=50$. (b)~Sparse Ising with fields. (c)~Weighted
 MaxCut. (d)~Weighted Max-2-XOR. Complete cells have 100 instances. Shaded
 regions with open symbols are the incomplete extension cells, excluded from every
 pooled statistic and conclusion: Ising $n=48$ (98) and $n=52$ (61),
 weighted MaxCut $n=52$ (55) and Max-2-XOR $n=52$ (46 of a designed 50);
 $\dagger$ marks cells censored by run time. Panel titles count all
 instances, including these cells. Bars are pointwise 95\%
 instance-bootstrap intervals. Classical selection work is additional to
 the one-mask angle optimization and is not equalized against pool
 construction.}
 \label{fig:one-vs-pool}
\end{figure*}

The stronger comparator is the random pool (Fig.~\ref{fig:one-vs-pool} and
Table~\ref{tab:benchmark-comparisons}). Across all MaxCut instances, MARS minus
Rand-SEMI is $-0.00256$ with interval $[-0.00535,0.00016]$; on the complete
Ising cohort it is $-0.0040\,[-0.0081,0.0001]$. Neither interval establishes
a difference, equivalence or noninferiority. On weighted MaxCut, Max-2-XOR
and Max-2-SAT the pooled difference favors the pool, by $0.0207$, $0.0240$
and $0.0061$, respectively.

The size dependence differs by objective. MaxCut and Ising show a
reversal within the measured sizes: the pointwise intervals
favor the pool at small sizes and MARS at larger ones, for example
$+0.0232\,[+0.0135,+0.0324]$ at MaxCut $n=48$ and
$+0.0240\,[+0.0159,+0.0318]$ at Ising $n=44$. For weighted MaxCut,
Max-2-XOR and Max-2-SAT the pool's advantage at small sizes becomes a
nonsignificant difference at the largest complete sizes, but no complete
size cell significantly favors MARS. These pointwise comparisons do not
establish a unique crossover size or justify extrapolation beyond the
measured sizes. One
selected mask therefore recovers substantial improvement over all-plus, but
does not replace the pool uniformly across objectives or sizes.

\begin{figure*}[t]
 \centering
 \includegraphics[width=\textwidth]{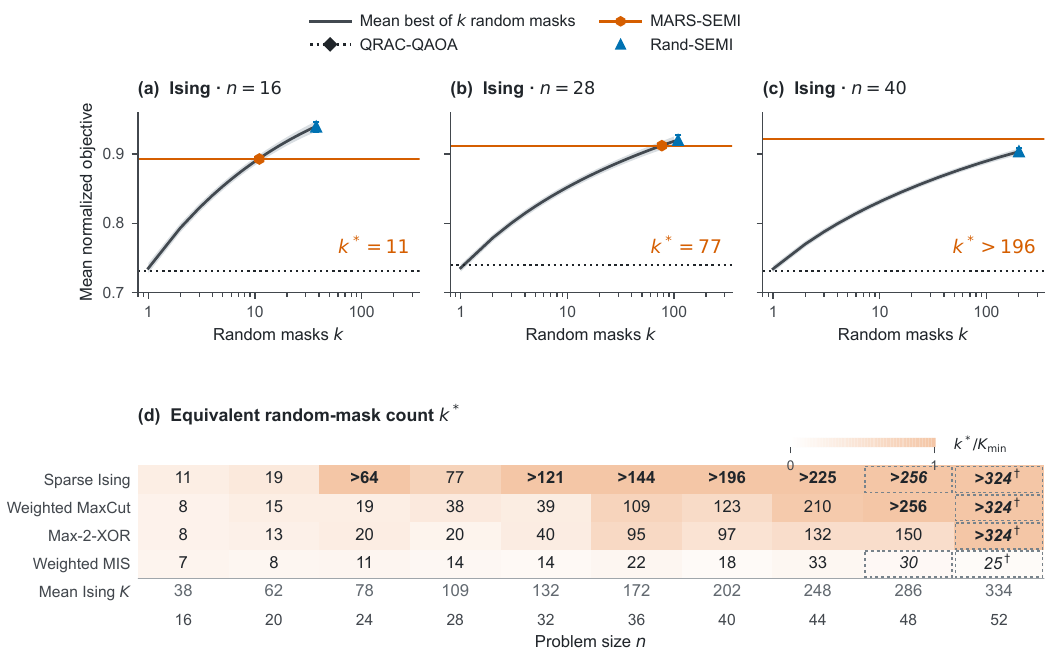}
 \caption{\textbf{How much random-mask search does one selected mask
 replace?} (a--c)~Complete Ising cohorts at $n=16$, 28 and 40, with
 100 instances each. Curves show the instance mean of
 Eq.~\eqref{eq:finite-pool} on $k\le K_{\min}$; grey bands are pointwise
 95\% instance-bootstrap intervals. Orange and dotted lines are MARS-SEMI
 and the all-plus baseline. Crossings are $11$, $77$, and right-censored
 beyond $196$, respectively. The blue triangle is the mean full-pool
 maximum, positioned at the mean pool size; varying individual pool sizes
 mean that it need not lie on the common-$k$ curve. (d)~$k^*$ in every
 size cell of the Ising, weighted MaxCut, Max-2-XOR and weighted
 independent-set (MIS, feasibility-aware score) families. Bold
 ``$>$'' entries are right-censored at $K_{\min}$; italic entries with
 dashed outlines are incomplete cells, and $\dagger$ marks cells censored
 by run time. Shading is $k^*/K_{\min}$ for observed crossings and saturates
 at one for right-censored entries; the bottom row gives the mean Ising pool size for
 orientation. The MaxCut and Max-2-SAT rows do not store per-mask pool
 values, so their $k^*$ is unavailable. Each $k^*$ is a point estimate
 without an interval. The comparison is conditional on the recorded pools
 and counts optimized masks, not quantum shots or equal total
 computational work.}
 \label{fig:equal-budget}
\end{figure*}

\subsection{Selection beyond one random mask}
The recorded pool values resolve quality as a function of the number of
angle-optimized random candidates, including the one-mask baseline.
For instance $g$, order the $K_g$ stored scores
as $v_{g,(1)}\le\cdots\le v_{g,(K_g)}$. The expected best of $k$ entries
drawn without replacement is exactly
\begin{equation}
 B_g(k)=\sum_{j=k}^{K_g}v_{g,(j)}
 \frac{\binom{j-1}{k-1}}{\binom{K_g}{k}},
 \qquad 1\le k\le K_g.
 \label{eq:finite-pool}
\end{equation}
This inherited finite-population accounting~\cite{semre2026randsemi}
includes tied values and requires no new statevector evaluations.
Within a size cell, average $B_g(k)$ over instances only on the common
domain $k\le K_{\min}:=\min_gK_g$. Let $k^*$ be the first $k$ where that
average reaches the mean MARS score; if none does, report $k^*>K_{\min}$.
Appendix~\ref{app:pool-statistics} gives the counting argument and the
distinction from the mean of the full-pool maxima.

Figure~\ref{fig:equal-budget}(a--c) isolates the role of selection on three
complete Ising cells. The crossing occurs at $k^*=11$ for $n=16$ and
$77$ for $n=28$. At $n=40$ there is no crossing for $k\le196$;
separately, MARS's mean $0.9216$ exceeds the full-pool mean $0.9033$.
By contrast, one random-pool candidate averaged over the 800 complete
Ising instances scores $0.7322$, close to the all-plus score $0.7319$.
The large one-mask improvement in this comparison is associated with
selection, not simply with drawing a nonconstant reference.
Figure~\ref{fig:equal-budget}(d) extends $k^*$ to every size cell of the
families whose rows store the per-mask pool values. In the complete cells
of Ising, weighted MaxCut and Max-2-XOR it grows from 8--11 at $n=16$ to
values between 132 and 210, or right-censored values, at $n=44$ and 48. For
independent set, where MARS trails the pool on the feasibility-aware score
(Appendix~\ref{app:feasibility}), it stays between 7 and 33.

These are exact summaries of finite stored pools, not guarantees over fresh
pool seeds or individual instances. Classical selection has its own cost,
and the stored timers do not separately record all of its construction and
initialization work.
They therefore do not support an end-to-end speedup comparison
(Appendix~\ref{app:protocol}).

\section{Discussion and conclusion}
\label{sec:discussion}
The main distinction is between the quantum states reachable by an ansatz
and the assignments reachable through its decoder. The cubic construction
shows that a physical symmetry can restrict the latter at every all-plus
depth. The weighted construction shows that a depth-one obstruction can
survive controlled coupling between modules. Both conclusions rely on
specific encodings and deterministic sign decoding. Changing the packing,
initialization axis or rounding rule changes the comparison.

The simulations establish a different, operational statement. Classical
selection can capture a large part of the benefit of mask variation with
one continuous-angle optimization: on the five unconstrained families, MARS
closes 58--69\% of the all-plus baseline's mean gap to the optimum, against
68--72\% for the $r^2$-mask pool (Appendix~\ref{app:max2sat}). Its
performance relative to the pool depends on size and objective. The pool
leads at small sizes in every family; MARS leads at the larger complete
MaxCut and Ising sizes, whereas for weighted MaxCut, Max-2-XOR and
Max-2-SAT the difference narrows to a nonsignificant one without an
established reversal.
Whether these trends continue beyond the measured sizes is open. The
finite-pool analysis supports a selection effect beyond drawing one random
mask. None of these observations identifies the constructed symmetry as the
mechanism on random instances.

Constraint feasibility further limits the conclusion. In the complete
100-instance independent-set cells at $n=40$ and 44, MARS returns a feasible
set on 50\% of instances, while the random pool returns one on all recorded
instances. Infeasible decodes receive zero and are not repaired
(Appendix~\ref{app:feasibility}). A favorable encoded objective alone does
not ensure a valid solution.

The resource claim is a reduction in the number of separately optimized
masks. It includes a classical selection task whose cost must be measured,
and it is not a quantum speedup. Noiseless statevector simulation, one frozen
pool seed per instance, and finite graph sizes leave hardware readout,
noise, seed variability and asymptotic performance untested. These limits
do not weaken the structural conclusion: initialization is an expressivity
choice that additional all-plus depth cannot always replace. They define
the scope in which the present selector results are evidence for making
that choice classically.

\section*{Data and code availability}
The simulation and selector code, campaign configurations, instance
manifests with certified optima, instance-level result rows, including the
stored per-mask Rand-SEMI values, and the source tables of the numerical figures are
held in the authors' project repository at revision \texttt{d493778}
(Appendix~\ref{app:protocol}). They are available from the authors on
reasonable request.

\section*{Author contributions}
Emilio Semre conducted the research and drafted the manuscript. Steven
Frankel supervised the work.

ChatGPT and Claude assisted with editing, mathematical checks, reference
verification, analysis of existing data, and figure and LaTeX preparation.
The authors take full responsibility for the work.
\begingroup
\interlinepenalty=10000
\setlength{\bibsep}{3pt}
\bibliographystyle{quantum}
\bibliography{references}
\endgroup
\label{page:main-end}
\onecolumn
\appendix
\section*{Technical appendices}
The appendices give the finite-angle estimates, complete proofs of the two
separation theorems and the motivating example, the protocol needed to
interpret the numerical comparisons, and two supplementary families:
weighted independent set and weighted Max-2-SAT. Only the specified encodings and decoder
are used in the proofs. The numerical studies use the separate instance
distributions in Appendix~\ref{app:protocol}.
\section{Local expansion and finite-angle control}
\label{app:local-bounds}

This appendix supplies the local estimates used in the two separation proofs.
Let $G=(V,E)$ be a finite simple unweighted graph of maximum degree
$\Delta\ge1$, with an admissible unsigned Pauli encoding: at most one logical
vertex occupies each axis of a physical qubit, and every logical edge joins
different qubits. Omitting the scalar part of the cost Hamiltonian, write
\begin{equation}
 H=-\frac32\sum_{ij\in E}P_iP_j,\qquad
 B_x=-\sum_qx_qX_q,\qquad
 \ket{\phi_x}=\bigotimes_q\ket{x_q}_X,
 \quad x\in\{\pm1\}^r.
 \label{eq:appendix-local-model}
\end{equation}
For $a\in\{X,Y,Z\}$, define the seed field
\begin{equation}
 F_{q,a}(x)=
 \sum_{\substack{j\sim(q,a)\\P_j=X_{q(j)}}}x_{q(j)},
 \label{eq:appendix-seed-field}
\end{equation}
with value zero when the slot $(q,a)$ is unoccupied. These are integers with
$|F_{q,a}|\le\Delta$. The depth-one marginal on axis $a$ is
\begin{equation}
 m_{q,a}=\bra{\phi_x}e^{i\gamma H}e^{i\beta B_x}
 \sigma_a^q e^{-i\beta B_x}e^{-i\gamma H}\ket{\phi_x}.
 \label{eq:appendix-marginal}
\end{equation}
Only occupied axes are decoded, using the common rule
$\operatorname{sign}_0(0)=+1$.

\begin{lemma}[Exact first-order terms with uniform remainder]
\label{lem:local-expansion}
Put $c=\cos(2\beta)$ and $t=\sin(2\beta)$. For every mask and all real
$\beta,\gamma$,
\begin{equation}
 \begin{aligned}
 m_{q,X}&=x_q+R_{q,X},\\
 m_{q,Y}&=3\gamma\bigl(tF_{q,Y}-cx_qF_{q,Z}\bigr)+R_{q,Y},\\
 m_{q,Z}&=3\gamma\bigl(tF_{q,Z}+cx_qF_{q,Y}\bigr)+R_{q,Z},
 \end{aligned}
 \label{eq:appendix-local-expansion}
\end{equation}
where
\begin{equation}
 |R_{q,X}|\le36\Delta^2\gamma^2,\qquad
 |R_{q,Y}|,|R_{q,Z}|
 \le36\Delta^2\gamma^2(|c|+|t|).
 \label{eq:appendix-local-error}
\end{equation}
\end{lemma}

\begin{proof}
The mixer pulls the measured Pauli back to
\begin{equation}
 O_X=X_q,\qquad O_Y=cY_q+x_qtZ_q,\qquad
 O_Z=cZ_q-x_qtY_q.
 \label{eq:appendix-mixer-pullback}
\end{equation}
Decompose the cost terms incident on $q$ as
$X_qh_X+Y_qh_Y+Z_qh_Z$, where $h_a$ acts away from $q$.
Product-state factorization gives
$\langle h_a\rangle=-3F_{q,a}/2$,
$\langle X_q\rangle=x_q$, and $\langle Y_q\rangle=\langle Z_q\rangle=0$.
The Pauli commutators therefore give
\begin{equation}
 \begin{aligned}
 i\langle[H,X_q]\rangle&=0,\\
 i\langle[H,Y_q]\rangle&=2x_q\langle h_Z\rangle=-3x_qF_{q,Z},\\
 i\langle[H,Z_q]\rangle&=-2x_q\langle h_Y\rangle=3x_qF_{q,Y}.
 \end{aligned}
 \label{eq:appendix-initial-derivatives}
\end{equation}
Combining these identities with Eq.~\eqref{eq:appendix-mixer-pullback}
proves the constant and linear terms in
Eq.~\eqref{eq:appendix-local-expansion}.

To bound the error for the exact evolution, set
$\mathcal L(O)=i[H,O]$ and $\alpha_u(O)=e^{iuH}Oe^{-iuH}$.
Taylor's integral remainder and unitary invariance of the operator norm give
\begin{equation}
 \left\|\alpha_\gamma(O)-O-\gamma\mathcal L(O)\right\|
 \le\frac{\gamma^2}{2}\|\mathcal L^2(O)\|.
 \label{eq:appendix-integral-remainder}
\end{equation}
Indeed, the second derivative is $\alpha_u(\mathcal L^2(O))$; integrating
its norm over the triangular remainder kernel yields $\gamma^2/2$.
For negative $\gamma$ the same argument reverses the integration direction.

Expand $O$ in Hermitian tensor-Pauli strings and bound its operator norm by
the sum of absolute coefficients. A string supported on $\ell$ qubits can
anticommute with at most $2\Delta\ell$ edge terms: at a supported qubit,
only its two other Pauli axes can contribute, each with at most $\Delta$
incident logical edges. Counting an edge twice only enlarges the bound.
Every nonzero commutator with $-3P_iP_j/2$ multiplies coefficient magnitude
by three and introduces at most one new qubit. Starting from coefficient
sum $\Gamma_0$, the first two commutators consequently have sums at most
$6\Delta\Gamma_0$ and $72\Delta^2\Gamma_0$. In
Eq.~\eqref{eq:appendix-mixer-pullback}, $\Gamma_0=1$ for $X$ and
$\Gamma_0=|c|+|t|$ for $Y,Z$. Substitution in
Eq.~\eqref{eq:appendix-integral-remainder}, followed by taking the initial
expectation, proves Eq.~\eqref{eq:appendix-local-error}.
No commutativity between different edge terms is assumed; multiple logical
edges between a pair of physical qubits are permitted.
\end{proof}

\subsection{The weighted cubic estimate used in the robust construction}
For a cubic graph with edge weights $0<w_{ij}\le1$, replace $H$ by
$-3\sum_{ij}w_{ij}P_iP_j/2$ and define
\begin{equation}
 F^w_{q,a}(x)=
 \sum_{\substack{j\sim(q,a)\\P_j=X_{q(j)}}}
 w_{(q,a),j}x_{q(j)}.
 \label{eq:weighted-seed-field}
\end{equation}
The same initial-derivative calculation applies with $F$ replaced by $F^w$.
At $\beta=\pi/4$ it reads
\begin{equation}
 m_{q,X}=x_q+R_{q,X},\qquad
 m_{q,Y}=3\gamma F^w_{q,Y}+R_{q,Y},\qquad
 m_{q,Z}=3\gamma F^w_{q,Z}+R_{q,Z}.
 \label{eq:weighted-local-expansion}
\end{equation}
We retain the following deliberately loose remainder estimate from the
weighted construction:
\begin{equation}
 |R_{q,a}|\le729\gamma^2\qquad(a=X,Y,Z).
 \label{eq:weighted-remainder}
\end{equation}
Here is its complete norm count. A physical qubit contains at most three
cubic logical vertices, so at most nine edge terms meet it. Each term has
norm at most $3/2$. At $\beta=\pi/4$, each pulled-back observable in
Eq.~\eqref{eq:appendix-mixer-pullback} is a single Pauli up to sign.
Its first commutator has coefficient sum at most $9\cdot3$ and support on
at most two qubits. At most eighteen edge terms can meet any resulting
string, and the next commutator multiplies coefficients by at most three.
Thus $\|\mathcal L^2(O_a)\|\le9\cdot18\cdot9=1458$.
Equation~\eqref{eq:appendix-integral-remainder} gives
Eq.~\eqref{eq:weighted-remainder}, uniformly in the graph size, mask and
cost angle. This estimate controls the exact finite-angle witness; it is
not a truncation of the circuit used to define its decoded score.

\section{Symmetry and the infinite cubic separation}
\label{app:cubic}

\subsection{The two-qubit example}
\label{app:k33}
We first verify Example~\ref{ex:k33}. If a physical unitary $U_g$ implements
a permutation of logical observables,
$U_gP_iU_g^\dagger=P_{g(i)}$, then expectation-sign decoding with a common
tie rule is equivariant: $D(U_g\rho U_g^\dagger)=gD(\rho)$, where
$(gs)_{g(i)}=s_i$. More generally, any deterministic decoder with this
property satisfies $D(\rho)=gD(\rho)$ whenever
$U_g\rho U_g^\dagger=\rho$. This follows by direct substitution and
requires no assumption that the marginals are nonzero.

An admissible two-qubit encoding of $K_{3,3}$ puts each bipartition on one
fully occupied qubit. Indeed, a qubit cannot contain vertices from both
parts because they are adjacent. With unsigned Pauli axes, every such
packing therefore has
\begin{equation}
 H=\frac92I-\frac32A_1A_2,
 \qquad A=X+Y+Z,\qquad B_+=-X_1-X_2.
 \label{eq:appendix-k33-cost}
\end{equation}
Qubit exchange commutes with both $H$ and $B_+$ and fixes
$\ket{++}_X$. Every all-plus output is consequently exchange-invariant
for every finite $p\ge1$ and every real angle sequence. A deterministic
exchange-equivariant decoder returning a labeled assignment must give
identical triples $(a,a)$ on the two parts. Its cut is
\begin{equation}
 C_{K_{3,3}}(a,a)=\frac12\left[9-\left(\sum_{j=1}^3a_j\right)^2\right]
 \in\{0,4\},\qquad a\in\{\pm1\}^3.
 \label{eq:appendix-k33-cap}
\end{equation}
This proves the all-depth upper bound, including exact-zero marginals.

For the attaining witnesses, put $N=A/\sqrt3$, so $N^2=I$. At
$\gamma=-\pi/18$, the cost unitary, up to global phase, is
$U_C=(I-iN\otimes N)/\sqrt2$. Starting from
$\rho_q=(I+x_qX)/2$, a partial trace over the other qubit $u\ne q$ gives
\begin{equation}
 \widetilde\rho_q=
 \frac12(\rho_q+N\rho_qN)
 -\frac{i x_u}{2\sqrt3}[N,\rho_q].
 \label{eq:appendix-k33-reduced}
\end{equation}
To check the cross term, use $\operatorname{tr}(N\rho_u)=x_u/\sqrt3$
in the product expansion of $U_C(\rho_1\otimes\rho_2)U_C^\dagger$.
The identities
\begin{equation}
 NXN=-\frac13X+\frac23(Y+Z),\qquad
 [N,\rho_q]=\frac{i x_q}{\sqrt3}(Y-Z)
 \label{eq:appendix-k33-paulis}
\end{equation}
then yield the cost-state Bloch vector
\begin{equation}
 \widetilde m_q=
 \frac{x_q}{3}\bigl(1,1+x_u,1-x_u\bigr).
 \label{eq:appendix-k33-cost-vector}
\end{equation}
Using the mixer pullbacks in Eq.~\eqref{eq:appendix-mixer-pullback}, at
$\beta=\pi/8$ the mask $(+1,-1)$ gives
\begin{equation}
 m_1=\frac13(1,\sqrt2,\sqrt2),\qquad m_2=-m_1.
 \label{eq:appendix-k33-witness}
\end{equation}
Every logical sign in one part is opposite to every sign in the other,
so all nine edges are cut, with minimum absolute marginal $1/3$.
At the same angles the all-plus mask gives
$m_1=m_2=(1,\sqrt2,-\sqrt2)/3$, which cuts four edges.
Both angles lie in $\mathcal D$. Hence the all-plus optimum over all finite
depths and real angles is four, while the depth-one SEMI optimum is nine.
The production encoder's two color classes give such a full packing.
No conclusion is asserted for encodings with more than two qubits.
The labeled-output condition is also essential: the optimal unoriented
partition is fixed by exchanging its two parts, and randomized decoding is
not constrained to return a fixed labeled assignment.

\subsection{Proof of the infinite-family theorem}
We prove Theorem~\ref{thm:cubic} on its stated encoding. For even $r\ge4$,
put $h=r/2$ and use indices modulo $r$. The three logical vertices
$X_q,Y_q,Z_q$ occupy their named axes of physical qubit $q$. The edge set is
\begin{equation}
 \begin{aligned}
 &\{X_q,X_{q+h}\},&&0\le q<h,\\
 &\{Y_q,Y_{q+1}\},\ \{Z_q,Z_{q+1}\},&&0\le q<r,\\
 &\{X_q,Y_{q+1}\},\ \{X_q,Z_{q+1}\},&&0\le q<r.
 \end{aligned}
 \label{eq:appendix-cubic-edges}
\end{equation}
Every $X$ vertex has one matching edge and two spokes; every transverse
vertex has two cycle edges and one spoke. Because $r\ge4$ and $h\ge2$,
these edges are distinct and join different physical qubits. The $Y$ cycle
is connected, and every $X$ and $Z$ vertex connects to it through the
spokes. Thus the graph is simple, connected and cubic, with $n=3r$
logical vertices, $9r/2$ edges and full compression $r=n/3$.

\paragraph{The all-depth obstruction.}
Let $U$ cyclically permute physical qubits, $q\mapsto q+1$, without
permuting their axes. It preserves each edge family in
Eq.~\eqref{eq:appendix-cubic-edges}, the all-plus mixer and the product
reference $\ket{+}^{\otimes r}$. It therefore fixes every all-plus output
state for all finite depths and all real angle vectors. For each
$a\in\{X,Y,Z\}$, the expectation of $\sigma_a^q$ is independent of $q$.
Applying the common deterministic decoder gives one sign on each axis
orbit, including when that common expectation is zero. All $XX$, $YY$
and $ZZ$ edges are uncut. There are only $2r$ remaining edges, so
\begin{equation}
 C_{\mathbf1,p}(\boldsymbol\beta,\boldsymbol\gamma)\le2r
 \quad\text{for every }p\ge1
 \text{ and }\boldsymbol\beta,\boldsymbol\gamma\in\mathbb R^p.
 \label{eq:appendix-cubic-all-depth}
\end{equation}

\paragraph{The exact finite-angle score.}
Each of $Y_q,Z_q$ has precisely one logical $X$ neighbor, namely $X_{q-1}$.
Thus $F_{q,Y}=F_{q,Z}=x_{q-1}$. At
\begin{equation}
 \theta_*=(\beta_*,\gamma_*)
 =\left(\frac12\arctan4,-\frac1{1080}\right)\in\mathcal D,
 \qquad (t,c)=\frac{(4,1)}{\sqrt{17}},
 \label{eq:appendix-cubic-angle}
\end{equation}
Lemma~\ref{lem:local-expansion} gives the transverse leading terms
\begin{equation}
 -\frac{(4-x_q)x_{q-1}}{360\sqrt{17}},\qquad
 -\frac{(4+x_q)x_{q-1}}{360\sqrt{17}}
 \label{eq:appendix-cubic-leading}
\end{equation}
for $Y_q,Z_q$, respectively. Both have sign $-x_{q-1}$ and magnitude
at least $3/(360\sqrt{17})$. Since $\Delta=3$, their errors are at most
\begin{equation}
 324\left(\frac1{1080}\right)^2\frac5{\sqrt{17}}
 =\frac1{720\sqrt{17}}.
 \label{eq:appendix-cubic-transverse-error}
\end{equation}
The exact transverse magnitudes are therefore at least
$1/(144\sqrt{17})>0$. Likewise,
$|m_{q,X}-x_q|\le324/1080^2=1/3600$, so every $X$ sign is $x_q$
with signed marginal at least $3599/3600$. Consequently every mask has
\begin{equation}
 \widehat s_{X_q}=x_q,\qquad
 \widehat s_{Y_q}=\widehat s_{Z_q}=-x_{q-1}.
 \label{eq:appendix-cubic-decoded}
\end{equation}
This is an exact nonzero-angle statement with strict margins on every
coordinate, rather than an inference from first derivatives alone.

Every spoke is cut in Eq.~\eqref{eq:appendix-cubic-decoded}. A cycle edge
on either transverse axis is cut precisely when $x_{q-1}\ne x_q$, and
the $X$ matching has the cut of the mask itself. With $C_r$ the physical-index
cycle and $M_h=\{\{q,q+h\}:0\le q<h\}$, the score is exactly
\begin{equation}
 f_r(x):=C_{x,1}(\theta_*)
 =2r+2\operatorname{Cut}_{C_r}(x)+\operatorname{Cut}_{M_h}(x).
 \label{eq:appendix-cubic-score}
\end{equation}
The all-plus mask attains $f_r(\mathbf1)=2r$, completing the equality in
the all-depth baseline bound.

\paragraph{Moments and the optimal subfamily.}
For a uniform mask, its coordinates are independent fair signs. Expanding
the cut indicators gives
\begin{equation}
 f_r(x)=\frac{13r}{4}
 -\sum_{q=0}^{r-1}x_qx_{q+1}
 -\frac12\sum_{q=0}^{h-1}x_qx_{q+h}.
 \label{eq:appendix-cubic-walsh}
\end{equation}
For $r\ge4$ the unordered pairs in both sums are distinct. Each nonconstant
sign monomial has expectation zero and square one. Two distinct monomials
have zero inner product, because some sign occurs to odd multiplicity in
their product. Hence
\begin{equation}
 \mathbb E f_r(X)=\frac{13r}{4},\qquad
 \operatorname{Var}f_r(X)=r+\frac h4=\frac{9r}{8}.
 \label{eq:appendix-cubic-moments}
\end{equation}
This calculation does not assume independence of the edge-cut indicators.
Subtracting the all-depth all-plus optimum gives the mean gap
$13r/4-2r=5r/4=5n/12$.

If $h$ is odd, the alternating mask $x_q=(-1)^q$ cuts every cycle edge
and every matching edge. Equation~\eqref{eq:appendix-cubic-score} then
equals the full edge count $9r/2$, which is a universal upper bound on
any cut. Thus $C^*_{\mathrm{SEMI},1}=\operatorname{OPT}(G_r)=9r/2$
and the gap from the all-depth all-plus optimum is $5r/2=5n/6$.
This proves all claims of Theorem~\ref{thm:cubic}. The encoding is part of
the construction; repacking can destroy its physical cyclic symmetry.

\section{Complete proof of the weighted separation}
\label{app:weighted}

This appendix proves Theorem~\ref{thm:weighted}. The comparison is at depth
one on $\calD=[0,\pi]\times[-\pi,63\pi/65]$. In fact, the all-plus upper
bound below holds for every real $\beta$ and every $|\gamma|\le\pi$.
The encoder ordering and the weak-edge interval are part of the hypotheses.

\subsection{The connected ensemble and its prescribed encoding}
\label{app:weighted-construction}

Fix $m\ge3$. For $t\in\{0,\ldots,m-1\}$, module $t$ has vertices
$L_{t,a},R_{t,a}$, where $a\in\{0,1,2\}$ denotes $X,Y,Z$, respectively.
Its six strong edges $L_{t,a}R_{t,b}$, $a\ne b$, have weight one. For
each axis, independently choose a uniform derangement $\pi_a$ of the
module indices and add the weak edges
\begin{equation}
 L_{t,a}R_{\pi_a(t),a},\qquad
 w_{t,a}\mathrel{\mathop{\sim}^{\mathrm{iid}}}
 U(1/400,1/200),
 \label{eq:weighted-weak-edges}
\end{equation}
with weights independent of the permutations. Condition this law on the
graph being connected; the weight interval is open, so every admissible
weight satisfies $1/400<w_{t,a}<1/200$. Write
\begin{equation}
 W_a=\sum_{t=0}^{m-1}w_{t,a},\qquad W=\sum_aW_a.
 \label{eq:weighted-layer-weights}
\end{equation}
Every realization is simple, bipartite and cubic: each vertex has two
strong neighbors on different axes and one weak neighbor on its own axis.
There are $n=6m$ vertices and $9m$ edges. The conditioning event has positive
probability for every $m\ge3$: choose all three permutations to be the same
$m$-cycle. Each strong module is connected, and these weak edges connect
the modules in a cycle. The separation proved below is pointwise for every
connected realization; it does not rely on a typical-sample estimate.

Assign distinct integer labels by
\begin{equation}
\begin{aligned}
 \ell(L_{t,a})&=3t+((a-t)\bmod3),\\
 \ell(R_{t,a})&=3m+3t+((a-(m+t))\bmod3).
\end{aligned}
\label{eq:weighted-labels}
\end{equation}
Apply the specified production encoder: find a minimum proper coloring
with ascending vertex order and ascending color trials; process the color
classes in increasing color order; sort each class by vertex label; and
pack consecutive triples, assigning position $j\in\{0,1,2\}$ of
zero-based qubit $q$ the axis $XYZ[(j+q)\bmod3]$. Vertex zero is $L_{0,0}$
and receives color zero. A connected bipartite graph has exactly two
proper two-colorings, so every $L$ vertex receives color zero and every
$R$ vertex receives color one. The sorted chunks are the module triples.
Equation~\eqref{eq:weighted-labels} therefore puts $L_{t,a}$ on axis $a$
of qubit $t$ and $R_{t,a}$ on axis $a$ of qubit $m+t$. This is an
admissible encoding with $r=2m=n/3$ fully occupied qubits. Ascending color
and vertex order are essential to this identification; the result does not
assert that an arbitrary relabeling or coloring implementation yields the
same packing.

\subsection{A nonvanishing coordinate in the uncoupled all-plus modules}
\label{app:weighted-bloch}

Remove the weak edges and omit scalar Hamiltonian terms. On one pair of
module qubits the cost generator is
\begin{equation}
 K_0=-\frac32\sum_{a\ne b}\sigma_a\otimes\sigma_b
     =-\frac92M\otimes M+3S-\frac32I,
 \label{eq:weighted-module}
\end{equation}
where $M=(X+Y+Z)/\sqrt3$ and $S$ exchanges the two qubits. The identity
follows from $\sum_a\sigma_a\otimes\sigma_a=2S-I$. The all-plus initial
state $\ket{+}_X\otimes\ket{+}_X$ lies in the symmetric subspace, which
is preserved by $M\otimes M$. Since $S$ is the identity there, the cost
evolution of this state is, up to a scalar phase, that generated by
$-9M\otimes M/2$.

Both $M\otimes I$ and $I\otimes M$ commute with this generator.
Their expectations thus remain $1/\sqrt3$. Exchange symmetry makes the
two post-cost Bloch vectors equal, say $\widetilde v$, and hence
\begin{equation}
 \|\widetilde v\|_2\ge
 \left|\widetilde v\cdot\frac{(1,1,1)}{\sqrt3}\right|
 =\frac1{\sqrt3}.
 \label{eq:weighted-bloch-norm}
\end{equation}
The all-plus mixer applies the same rotation about $X$ to both vectors.
Consequently the final unperturbed Bloch vector $v^0(\beta,\gamma)$ is
the same on both qubits of every module and satisfies
\begin{equation}
 \|v^0(\beta,\gamma)\|_2\ge\frac1{\sqrt3},\qquad
 \max_a|v^0_a(\beta,\gamma)|\ge\frac13.
 \label{eq:weighted-nonvanishing}
\end{equation}
For each angle pair, choose an axis attaining this maximum. The chosen axis
may depend on the angles; it is common to all modules at those angles.

\subsection{A module-local perturbation bound and the all-plus cap}
\label{app:weighted-duhamel}

Let $K^{(0)}$ be the sum of the uncoupled module generators and $V$ the
weak-edge generator, so the full scalar-free cost is $K=K^{(0)}+V$.
With $\tau_u^A(O)=e^{iuA}Oe^{-iuA}$, differentiating
$\tau_s^K(\tau_{\gamma-s}^{K^{(0)}}(O))$ and integrating gives
\begin{equation}
\begin{split}
 \tau_\gamma^K(O)-\tau_\gamma^{K^{(0)}}(O)
 ={}&i\int_0^\gamma\tau_s^K\bigl(
 [V,\tau_{\gamma-s}^{K^{(0)}}(O)]\bigr)\,ds.
\end{split}
\label{eq:weighted-duhamel}
\end{equation}
This identity also holds for negative $\gamma$, with the oriented
integral. If $O$ is supported in one module, its unperturbed evolution
stays in that module. Exactly six weak edges meet the module: three
outgoing from its $L$ vertices and three incoming to its $R$ vertices.
They are distinct because the permutations are derangements. Every such
Hamiltonian term has norm $3w_{t,a}/2<3/400$, and all other terms commute
with the inner observable. Unitary conjugation preserves operator norm,
so for $\|O\|\le1$,
\begin{equation}
 \|\tau_\gamma^K(O)-\tau_\gamma^{K^{(0)}}(O)\|
 \le2\cdot6\cdot\frac3{400}|\gamma|
 =\frac{9}{100}|\gamma|.
 \label{eq:weighted-local-perturbation}
\end{equation}
Only the inner, uncoupled evolution is confined to the module; no locality
assumption is imposed on the outer interacting evolution. This is why the
bound is independent of the total number of modules despite the extensive
norm of $V$.

Pulling a measured Pauli observable back through the all-plus mixer leaves
its norm equal to one and its support in its original module. Thus
Eq.~\eqref{eq:weighted-local-perturbation} bounds the change of every
final marginal. For $|\gamma|\le\pi$, the change is at most
$9\pi/100<1/3$. Choose the common axis $a$ from
Eq.~\eqref{eq:weighted-nonvanishing}. Every actual marginal on that axis,
on every $L$ and $R$ qubit, has the same strict sign as $v^0_a$; its
absolute value is at least
\begin{equation}
 \frac13-\frac{9\pi}{100}>\frac{53}{1050}>0.
 \label{eq:weighted-common-sign-margin}
\end{equation}
The rational inequality follows from $\pi<22/7$.

Within module $t$, the vertices $L_{t,a}$ and $R_{t,a}$ therefore have
equal decoded signs. The six strong edges form a cycle, with two
edge-disjoint length-three paths between these vertices. Cutting every
edge of an odd-length path would require opposite endpoint signs.
Each path has an uncut edge, so at most four strong edges in that module
are cut. Every weak edge on axis $a$ is also uncut, since its endpoints
share the same strict sign across modules. The remaining weak edges can
contribute at most $W-W_a$. For the sign decoder, including its prescribed
tie rule on the other axes, this proves pointwise in the angles
\begin{equation}
 C_{\allplus,1}(\beta,\gamma)
 \le4m+W-W_a\le4m+W-\min_bW_b,
 \quad |\gamma|\le\pi.
 \label{eq:weighted-allplus-cap}
\end{equation}
In particular, the same cap holds for $C^*_{+,1}$ on $\calD$.

\subsection{A finite-angle mask recovering the entire bipartition}
\label{app:weighted-witness}

Assign $x_t=+1$ and $x_{m+t}=-1$ for every module, and choose
\begin{equation}
 \beta=\pi/4,\qquad\gamma=-1/486.
 \label{eq:weighted-witness-angles}
\end{equation}
These angles lie in $\calD$. Each $Y$- or $Z$-axis logical vertex has
exactly one strong $X$-axis neighbor, in the other half of its own
module, whose mask sign is $-x_q$. Its other strong neighbor and its weak
neighbor are transverse. Equations~\eqref{eq:weighted-local-expansion}
and~\eqref{eq:weighted-remainder} of Appendix~\ref{app:local-bounds}
consequently give
\begin{equation}
 \begin{aligned}
 m_{qX}&=x_q+R_{qX},\\
 m_{qY}&=-3\gamma x_q+R_{qY},\\
 m_{qZ}&=-3\gamma x_q+R_{qZ},
 \end{aligned}
 \qquad |R_{qa}|\le729\gamma^2.
 \label{eq:weighted-witness-expansion}
\end{equation}
The remainder estimate uses only cubic degree and weights at most one;
its complete double-commutator count is given in
Appendix~\ref{app:local-bounds}. It controls the exact circuit, including
noncommuting terms and repeated physical-qubit pairs.

Substitution of Eq.~\eqref{eq:weighted-witness-angles} yields
\begin{equation}
 \begin{aligned}
 x_qm_{qY},\ x_qm_{qZ}
 &\ge\frac3{486}-\frac{729}{486^2}=\frac1{324},\\
 x_qm_{qX}&\ge1-\frac{729}{486^2}=\frac{323}{324}.
 \end{aligned}
 \label{eq:weighted-witness-margins}
\end{equation}
All three logical variables on each qubit therefore decode to $x_q$
strictly. Every $L$ vertex is decoded $+1$ and every $R$ vertex $-1$,
cutting every edge. Positivity of the edge weights then proves
\begin{equation}
 C^*_{\SEMI,1}=\OPT=6m+W.
 \label{eq:weighted-optimum}
\end{equation}
Together with Eq.~\eqref{eq:weighted-allplus-cap}, this gives
\begin{equation}
 \Delta:=C^*_{\SEMI,1}-C^*_{+,1}
 \ge2m+\min_aW_a>(2+1/400)m.
 \label{eq:weighted-appendix-gap}
\end{equation}
The final strict inequality uses $W_a>m/400$ for every axis. This
completes the proof of Theorem~\ref{thm:weighted}. The upper bound is
restricted to depth one and the stated cost-angle interval; no
arbitrary-depth extension or different weak-edge distribution is used.
\hfill$\square$

\section{Experimental protocol and statistical definitions}
\label{app:protocol}

\subsection{Instances and normalization}

The comparisons use the cohorts in Table~\ref{tab:focused-cohorts}. Each method is evaluated on the same instances within a comparison. The retained MaxCut benchmark contains 1,699 of its 1,700 designed instances: one task failed at $n=50$, whose 99 successful instances remain in that benchmark. This exception is disclosed rather than described as a complete cell. Every other listed size cell is complete. The additional incomplete cells, $n=52$ of weighted MaxCut (55 of 100), Ising (61), Max-2-XOR (46 of a designed 50), Max-2-SAT (64) and independent set (56), and $n=48$ of Ising (98) and independent set (99), are excluded from every pooled statistic. Their campaigns were stopped by walltime limits or by decision before these cells were complete. All simulations use full, noiseless statevectors with floating-point numerical evolution and expectation values. They are classical numerical simulations, and their outputs are not exact-arithmetic certificates. The MaxCut cells $n=18,22,26,30,34$ used a dense-matrix statevector backend and all other cells a Pauli-term statevector backend; each size cell, and hence every paired comparison, uses a single backend.

\begin{table}[htbp]
\centering
\small
\caption{Included benchmark cohorts, all at $p=1$. $r$ is the number of physical qubits and $K=r^2$ the Rand-SEMI pool size. There are 100 instances per size, except the retained 99 of 100 for MaxCut at $n=50$.}
\label{tab:focused-cohorts}
\begin{tabular}{@{}llrrr@{}}
\toprule
Problem & Sizes $n$ & Instances & $r$ & $K$\\
\midrule
MaxCut & $18,20,\ldots,50$ & 1,699 & $6$--$18$ & $36$--$324$\\
Weighted MaxCut & $16,20,\ldots,48$ & 900 & $6$--$18$ & $36$--$324$\\
Sparse Ising & $16,20,\ldots,44$ & 800 & $6$--$16$ & $36$--$256$\\
Weighted Max-2-XOR & $16,20,\ldots,48$ & 900 & $6$--$18$ & $36$--$324$\\
Weighted Max-2-SAT & $16,20,\ldots,48$ & 900 & $6$--$18$ & $36$--$324$\\
Weighted independent set & $16,20,\ldots,44$ & 800 & $6$--$16$ & $36$--$256$\\
\bottomrule
\end{tabular}
\end{table}

\paragraph{MaxCut and weighted MaxCut.}
The graphs are seeded random connected 3-regular graphs: instance index $g$ uses the first of the seeds $7000+100g,7000+100g+1,\ldots$ for which the NetworkX sample is connected. The weighted instances reuse the corresponding unweighted topologies and assign edge weights uniformly from $\{1,\ldots,200\}/100$, using coefficient seed $20260912+1000n+g$. The decoded cut is divided by its certified classical maximum. Distinct objectives on reused topologies are not independent graph ensembles. The 300 MaxCut graph instances at $n=18,20,22$ belong to the scale cohort of the earlier Rand-SEMI study~\cite{semre2026randsemi}, which used the same generator, and the weighted MaxCut, Ising and independent-set instances at $n=20$ reuse those topologies; these are not an independent graph replication of that benchmark. The comparison with the classical selectors is the empirical contribution here.

\paragraph{Sparse Ising.}
The same cubic graph collection supports
\begin{equation}
 F(s)=\sum_i h_i s_i+\sum_{\{i,j\}\in E}J_{ij}s_i s_j,
 \qquad
 R_I=\frac{F(\widehat s)-F_{\min}}{F_{\max}-F_{\min}}.
 \label{eq:protocol-ising}
\end{equation}
The couplings are sampled uniformly from $\{\pm2,\ldots,\pm8\}/4$ and the fields from $\{\pm1,\ldots,\pm6\}/4$, with coefficient seed base $20261020$. Both classical extrema are certified, and their difference is nonzero on these instances. $R_I$ is a range-normalized objective, not a multiplicative approximation ratio. For dynamics, the encoded Hamiltonian is divided by $a=\max\{1,2\max_i|h_i|,2\max_{\{i,j\}\in E}|J_{ij}|\}$; decoded values use the original objective. This rescaling is part of the angle protocol.

\paragraph{Weighted Max-2-XOR.}
The clause graph for instance index $g$ is the NetworkX $G(n,m)$ sample with $m=3n/2$ edges and seed $7000+100g$, with one weighted parity clause $x_i\oplus x_j=b$ per edge. In sorted edge order, the parities $b$ are drawn uniformly from $\{0,1\}$ and then the weights uniformly from $\{1,\ldots,200\}/100$, using seed $20260917+1000n+g$. Quality is satisfied weight divided by its certified maximum. These graphs have a fixed edge count and need not be regular or connected.

\subsection{Encoding, mask selection, and variational budgets}

The maintained implementation uses the $(3,1)$-QRAC encoder \encoder. It finds a minimum proper coloring by deterministic backtracking, orders each color class by vertex label, partitions each class into consecutive groups of at most three, and assigns slot $j$ of physical qubit $q$ the axis \texttt{XYZ[(j+q)\%3]}, with zero-based indices. All compared masks share this frozen encoding. The decoder uses $\sgnzero(t)=+1$ for $t\geq0$ and $-1$ otherwise. Expectations close to zero can therefore be sensitive to numerical precision.

Each mask receives a depth-one angle search with 33 mixer-angle grid points and 65 cost-angle grid points, followed by L-BFGS-B refinement~\cite{byrd1995} from the three best grid candidates, with at most 80 iterations per start. Grid candidates are ordered by decreasing energy, then increasing $\beta$ and $\gamma$; a refinement replaces the incumbent only for strictly larger energy. The refinement bounds are the grid extrema, with \texttt{ftol}$=10^{-12}$, \texttt{gtol}$=10^{-5}$ and \texttt{maxfun}$=15{,}000$ per start. The optimized quantity is encoded energy; the reported quantity is the objective of the final sign-decoded assignment. Local angle optimization does not certify a global angle optimum.

The QRAC baseline uses the all-plus $X$ product state and its matching mixer. FAS-SEMI chooses one mask using the first-order \texttt{D1} surrogate with deterministic \texttt{sign0} ties; every production row uses this convention. ARES-SEMI refines a mask using the Adaptive-RLC2 surrogate, and MARS-SEMI searches that surrogate with an ARES warm start. Their recorded reference angles are $(\beta,\gamma)=(0.418,-0.127)$, after which the selected mask receives its own angle optimization. A warm start alone does not establish a retained-incumbent guarantee. No theorem requiring certified surrogate optimization or a finite-angle sign bound is applied to these implementation outputs without its premises.

Rand-SEMI optimizes $K=r^2$ distinct masks, drawn uniformly without replacement from the $2^r$ masks with a deterministic per-instance seed derived from master seed $20260817$ (Max-2-SAT: $20260830$), and retains the best decoded objective. The normalization by a certified optimum is unnecessary for this within-instance ranking. The candidate-count ranges are listed in Table~\ref{tab:focused-cohorts}. The MaxCut mean is 163.1 masks per instance; its per-size means range from 50.6 to 315.5. These averages are not individual pool sizes.

One selected mask therefore saves repeated variational optimizations relative to a pool; it does not mean one circuit call. Surrogate construction and search, grid evaluations, local optimization, and any physical measurement repetitions remain separate costs. The experiments do not measure shot budgets or a quantum-computational advantage, and do not establish polynomial-time mask selection.

The stored CPU timers do not provide a complete MARS pipeline cost.
The task residual subtracts the recorded RLC2 build, selector
searches, and the four nonpool angle optimizations from whole-task time;
it is not directly instrumented pool-only time. The recorded MARS components are
\texttt{rlc2\_build\_seconds}, \texttt{ares\_semi\_select\_seconds},
\texttt{mars\_semi\_select\_seconds} and
\texttt{mars\_semi\_qaoa\_seconds}. In the implementation the ARES timer
starts after its CEDAR initialization, and first-order surrogate
construction has no separate timer. These omitted costs, together with
encoding and bookkeeping, remain in the residual rather than the recorded MARS
components. Consequently these timings do not establish end-to-end
speedups. They are classical simulation
timings on the recorded cluster hardware.

\subsection{Classical selector specification}
\label{app:selectors}

The following specifies the frozen selector design underlying the rows;
all predictions use $(\beta_0,\gamma_0)=(0.418,-0.127)$, before the separate
angle optimization. Write $H$ for the encoded dynamics Hamiltonian and
$\overline P_i=e^{i\beta_0B_x}P_i e^{-i\beta_0B_x}$.
For a general quadratic objective the first-order field is
\begin{equation}
 L_i(x)=\langle\overline P_i\rangle_x+
 i\gamma_0\langle[H,\overline P_i]\rangle_x,
 \qquad D_1(x)=F_{\rm sel}\bigl(\sgnzero L(x)\bigr).
\end{equation}
Only tensor products of $I$ and $X$ survive the expectation in
$\ket{\phi_x}$, so $L_i$ is a local multilinear polynomial in mask signs.
For MaxCut this reduces to Eq.~\eqref{eq:selector-predictor}, with weighted
neighbor sums for weighted MaxCut. Write the selector objective as
$F_{\rm sel}(s)=c_0+\sum_i a_i s_i+\sum_{i<j}b_{ij}s_i s_j$.
It is the original quadratic objective, including the penalty quadratic for
independent set; its final evaluation still uses the native
feasibility-aware metric.

\paragraph{Adaptive-RLC2 predictor.}
Identical physical Pauli terms are first combined and zero terms discarded.
For a term subset $T$, let
$M_i(T;x)=\langle e^{i\gamma_0H_T}\overline P_i
e^{-i\gamma_0H_T}\rangle_x$, evaluated by diagonalizing the Hamiltonian
on its local support. If $B_q$ is the induced term set of the base cluster
rooted at $q=q(i)$, the predictor is
\begin{equation}
 \begin{aligned}
 \widehat m_i(x)={}&M_i(B_q;x)
 +\sum_{\substack{e\ni q\\e\notin B_q}}
   [M_i(\{e\};x)-M_i(\varnothing;x)]\\
 &+\sum_{\substack{\{e,f\}\ \text{rooted, connected}\\
                    \{e,f\}\not\subseteq B_q}}
   [M_i(\{e,f\};x)-M_i(\{e\};x)-M_i(\{f\};x)+M_i(\varnothing;x)].
 \end{aligned}
 \label{eq:rlc2-operational}
\end{equation}
A rooted connected pair has at least one term incident on $q$, with the
other sharing a qubit with it; unordered pairs are counted once. Each
bracket is tabulated over all mask signs on its support. Each predicted
marginal is a sum of local table lookups; applying $\sgnzero$ and
$F_{\rm sel}$ gives
$D_{\rm RLC2}(x)=F_{\rm sel}(\sgnzero\widehat m(x))$.
The base is evolved at finite $\gamma_0$ rather than
Taylor-truncated; singleton and pair corrections retain the corresponding
linked contributions outside it.

The base begins at $\{q\}$, including induced one-body terms, and grows by
one neighboring physical qubit at a time up to
$K_{\max}=\min\{r,4+\lceil\log_2\log_2(r+2)\rceil\}$.
Growth stops earlier if the frontier is empty or if every boundary term
commutes with every Pauli word generated from the root observables by
repeated nonzero commutators with internal terms. An occupied $Y$ or $Z$
slot includes both axes in this closure test. Candidate qubits are ranked
by finite-angle response per added cost, with cost proxy
$\widehat C(S)=8^{|S|}+|E[S]|4^{|S|}+|I(q)|2^{|S|}$,
where $E[S]$ is the set of physical Hamiltonian terms supported inside
$S$ and $I(q)$ the set incident on the root.
Specifically, for each newly internal term $h_e$, form the derivative
$\mathcal R_{T,e}(O)=\partial_\lambda
[e^{i\gamma_0(H_T+\lambda h_e)}Oe^{-i\gamma_0(H_T+\lambda h_e)}]_{\lambda=0}$.
Priority levels zero, one and two are, respectively, the
Pauli-coefficient $\ell_1$ norms of
$\mathcal R_{B_q,e}$,
$\mathcal R_{B_q,e}-\mathcal R_{\varnothing,e}$, and
$\mathcal R_{B_q,e}-\mathcal R_{\varnothing,e}
-\sum_{f\in B_q}(\mathcal R_{\{f\},e}-\mathcal R_{\varnothing,e})$.
Norms are summed over new terms, maximized over the two root mixer signs,
and summed over root observables with weights equal to incident cut weight
(MaxCut) or $|a_i|+\sum_j|b_{ij}|$ (general quadratics).
The lexicographic ranking uses levels two, one, zero, then
$\sum_e(e^{2|\gamma_0 h_e|}-1)$, each divided by the added cost, and finally
the smaller vertex label. Here $h_e$ in the scalar exponential denotes the
term coefficient. This deterministic rule and Eq.~\eqref{eq:rlc2-operational}
specify the local factor tables used by the surrogate.

\paragraph{Search and solver arithmetic.}
FAS first forms a deterministic CEDAR incumbent: fix mask bits by
conditional expectation of $D_1^2$ for nonnegative MaxCut scores, or of
$D_1$ for general signed quadratics, compare the seed against all-plus,
then take best improving one-bit or factor-co-occurring two-bit flips
until none remains. Bit order is decreasing factor-incidence count then
increasing index (the $D_1^2$ order also counts intersecting factor pairs);
branch ties prefer the larger conditional mean and then $+1$.
The seed comparison favors all-plus on a tie; move ties favor the first
move in index order. FAS compiles each $D_1$ sign into a Boolean
allowed-assignment table and maximizes its predicted objective by CP-SAT,
using the incumbent as hint, lower bound and fallback.

ARES compares the CEDAR mask and all-plus by $D_{\rm RLC2}$, keeps
all-plus on a tie, and applies strict best-improvement one-bit and
factor-co-occurring two-bit flips until no improving move exists.
MARS compiles the same RLC2 tables, rounding each factor entry at scale
$10^9$ to the nearest integer (ties to even). A negative integer field
sum predicts $-1$; zero predicts $+1$. MaxCut objectives use integer
weight ticks; general quadratic coefficients are rounded at scale $10^9$
(reduced if needed to keep integer activities below $2^{55}$).
The ARES mask supplies MARS's hint, lower bound and fallback in this
quantized arithmetic. The general-quadratic implementation additionally
returns the incumbent if its floating-score comparison worsens; the
MaxCut form guarantees only the quantized surrogate bound. Neither
guarantees the decoded score after angle optimization.

FAS and MARS use one CP-SAT worker and a 600-second limit per solve.
The exception is the small Max-2-SAT cohort, which uses 60 seconds with an
exhaustive surrogate cross-check. The remaining campaign configurations
require an \texttt{OPTIMAL} solver status; where enabled, exhaustive
cross-checks are restricted to $r\le14$. Such status certifies the
compiled surrogate objective, not the true quantum decoded objective,
and does not imply polynomial-time search.

\subsection{Uncertainty and finite-pool comparisons}
\label{app:pool-statistics}

Means weight each included instance once. Method differences are paired before averaging. The analysis uses instance-level percentile bootstrap intervals~\cite{efron1979} with 10,000 resamples and seed 12. Wins, ties, and losses use tolerance $10^{-12}$. Intervals are pointwise, without a simultaneous-coverage or equivalence claim. Bootstrap resampling carries an instance's complete mask pool together; masks are not additional independent problem instances. These intervals do not include systematic numerical error or uncertainty from algorithm development.

Equation~\eqref{eq:finite-pool} gives the expected best of $k$ candidate
indices sampled uniformly without replacement from each stored pool. Its
combinatorial factor is the probability that the largest selected index is
$j$, so the identity also holds for repeated score values. For $N$ instances,
define $\overline B(k)=N^{-1}\sum_g B_g(k)$ on the common domain
$k\le K_{\min}=\min_g K_g$. If $\overline M$ is the mean MARS-SEMI score,
the crossing index is
\begin{equation}
 k^*=\min\{k\le K_{\min}:\overline B(k)\ge\overline M\}.
\end{equation}
An empty set gives right-censoring beyond $K_{\min}$. This alone does not
establish a comparison with the mean full-pool score
$N^{-1}\sum_g B_g(K_g)$ when pool sizes differ.
The full-pool triangles in Fig.~\ref{fig:equal-budget} are plotted separately at mean pool size. The calculation is exact conditional on the recorded pools; it is not a population-level random-search guarantee.

\subsection{Reproducibility record}

Every empirical number and numerical figure in this paper is computed from
the instance-level campaign rows held in the authors' project repository
at revision \texttt{d493778}. The numerical figures have been redrawn from
the committed source tables, preserving the plotted values and intervals;
the conceptual illustration adds no numerical evidence. The quoted
statistics were recomputed from the datasets with the conventions above.
Fixed-seed bootstrap endpoints can depend on row order at the fourth
decimal; figure curves retain their source-table ordering, and quoted
dataset statistics use canonical order. Each campaign directory holds its
configuration and instance manifest, and its rows record the encoder
version and, for every method, the mask, optimized angles, relaxed energy
and decoded value; the weighted MaxCut, Ising, Max-2-XOR and independent-set
rows also store every per-mask Rand-SEMI value used in
Eq.~\eqref{eq:finite-pool}.

Appendix~\ref{app:selectors} states the operational selector definitions;
the source code and campaign configurations at that revision retain the
implementation and platform details. The angle grid takes 33 equally
spaced values of $\beta\in[0,\pi]$, including both endpoints, and 65 values
of $\gamma\in[-\pi,\pi)$, whose largest element $63\pi/65$ fixes the domain
$\mathcal D$ of Eq.~\eqref{eq:domain}. The Max-2-SAT rows were produced by
the campaign code that preceded a restructuring of the repository;
Appendix~\ref{app:max2sat} states how they were validated.

\clearpage
\setcounter{figure}{0}
\renewcommand{\thefigure}{S\arabic{figure}}
\renewcommand{\theHfigure}{supp.\arabic{figure}}
\section{A limitation for constrained objectives}
\label{app:feasibility}

Weighted maximum independent set illustrates a limitation of applying unconstrained mask selectors to a penalty encoding. The instances use random cubic graphs, positive vertex weights uniform in $\{1,\ldots,200\}/100$, and coefficient seed base $20260929$. The circuit optimizes an encoded penalty QUBO, but the evaluated score is the original constrained objective. If $I$ is the decoded set and $W^*$ its certified optimum, the reported quality is
\begin{equation}
 R_{\mathrm{IS}}(I)=
 \begin{cases}
 w(I)/W^*,& I\text{ is independent},\\
 0,&\text{otherwise}.
 \end{cases}
\end{equation}
No repair is applied. In integer weight units, the encoded objective is
$Q(x)=\sum_iw_ix_i-P\sum_{ij\in E}x_ix_j$ with $P=\max_iw_i+1$.
This sufficient penalty makes every maximizer a maximum-weight independent
set: removing a selected vertex incident on a violation improves $Q$ by at
least $P-w_i>0$. It is not claimed to be the minimum sufficient penalty
for each individual graph. The classical selectors also score this penalty
objective, whereas pool winners and final comparisons use the
feasibility-aware native score above.

The complete size cells already establish the practical limitation. At $n=16$, MARS-SEMI decodes a feasible set on 76 of 100 instances and achieves mean feasibility-aware ratio $0.5940$, whereas Rand-SEMI is feasible on all 100 and achieves $0.8212$. At $n=40$, these values are respectively 50 of 100 and $0.3972$, compared with 100 of 100 and $0.7253$; at $n=44$, the largest complete cell, they are 50 of 100 and $0.3998$, compared with 100 of 100 and $0.6838$. Thus a favorable objective value conditional on feasibility does not imply a favorable unconditional algorithmic output. Conditional means in Fig.~\ref{fig:mis-feasibility}(b) use different subsets for different methods and do not establish a paired superiority claim. Rand-SEMI's observed feasibility on every instance is an empirical outcome of retaining the best member of each pool, not a feasibility guarantee.

\noindent\begin{minipage}{\textwidth}
\centering
\includegraphics[width=\textwidth]{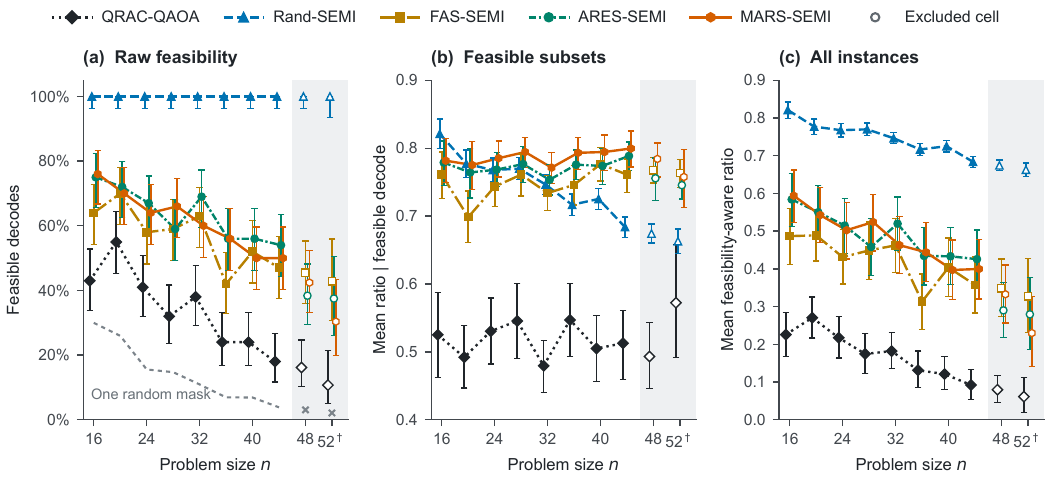}
\captionsetup{hypcap=false}
\captionof{figure}{\textbf{Feasibility limits mask selection for weighted independent set.}
Noiseless $p=1$ simulations on cubic graphs, with 100 instances at each of $n=16,20,\ldots,44$. Open symbols mark the incomplete cells at $n=48$ (99 instances) and $n=52$ (56), and $\dagger$ marks the run-time-censored $n=52$ cell; the conclusions in the text use the complete cells.
(a)~Fraction of decoded sets that are independent, with Wilson 95\% intervals~\cite{wilson1927}. The grey dashed line averages the feasible fraction within each stored random-mask pool.
(b)~Mean objective ratio conditional on feasibility, with algorithm-dependent feasible subsets.
(c)~Unconditional feasibility-aware ratio, assigning zero to infeasible outputs. Error bars in (b,c) are pointwise 95\% instance-bootstrap intervals. All methods use the same fixed encoding and sign decoder; decoded assignments are never repaired. Rand-SEMI optimizes $r^2$ masks, whereas each other method optimizes one.}
\label{fig:mis-feasibility}
\end{minipage}

\clearpage
\section{Weighted Max-2-SAT}
\label{app:max2sat}

The frozen weighted Max-2-SAT family places two clauses on each edge of a seeded connected cubic graph, hence $3n$ clauses. Literal signs are uniform apart from four fixed literal patterns that are always present, and clause weights are uniform in $\{4,\ldots,8\}/4$. The generator applies a fixed anti-cancellation repair and accepts only instances whose optimum is below the total weight. The instance seed is $20260830+300000+1000n+g$ for instance index $g$. The score is the satisfied clause weight divided by the certified optimum, and the dynamics use the objective divided by $\max\{1,2\max|\text{coefficient}|\}$. Encoding, decoder and per-mask optimizer are those of Appendix~\ref{app:protocol}.

These rows differ from the main cohorts in four disclosed respects. They were produced by the campaign code that preceded a restructuring of the repository; its FAS-, ARES- and MARS-SEMI masks coincide with those of the maintained implementation on its four reference instances ($n=16$--$28$). The Rand-SEMI pools use master seed $20260830$. The per-mask pool values were not stored, so Eq.~\eqref{eq:finite-pool} and $k^*$ are unavailable. Finally, the configuration of this family was validated for $n\le24$; at $n=28$ it failed a preregistered surrogate-accuracy check, a fixed-mask score correlation of $0.9732$ against the threshold $0.98$, and the sizes $n\ge28$ were run deliberately as a scaling diagnostic. The $n=52$ cell is incomplete, with 64 of 100 instances; 35 of the 36 missing instances have $r=19$, so the cell is censored by size and excluded.

On the 900 complete instances, $n=16,20,\ldots,48$, the mean ratios are $0.9055$ for all-plus, $0.9678$ for MARS and $0.9739$ for Rand-SEMI. The paired MARS improvement over all-plus is $+0.0623\,[+0.0599,+0.0647]$, with 861 wins, 5 ties and 34 losses, and MARS minus Rand-SEMI is $-0.0061\,[-0.0073,-0.0049]$. Per size, the pool is ahead at $n=16$--36 and the difference is not significant at $n=40$--48 (Fig.~\ref{fig:max2sat}b). The positive value $+0.0064\,[+0.0026,+0.0100]$ at $n=52$ comes from the censored cell and is not evidence of a crossover.

Because all-plus already scores $0.906$, the absolute improvements are smaller than in the other families. Figure~\ref{fig:max2sat}(c) therefore normalizes them by the all-plus gap to the optimum. MARS closes $0.659\,[0.645,0.674]$ of that gap and the pool $0.724\,[0.714,0.734]$, within the ranges $0.58$--$0.69$ and $0.68$--$0.72$ spanned by the five unconstrained families.

\noindent\begin{minipage}{\textwidth}
\centering
\includegraphics[width=\textwidth]{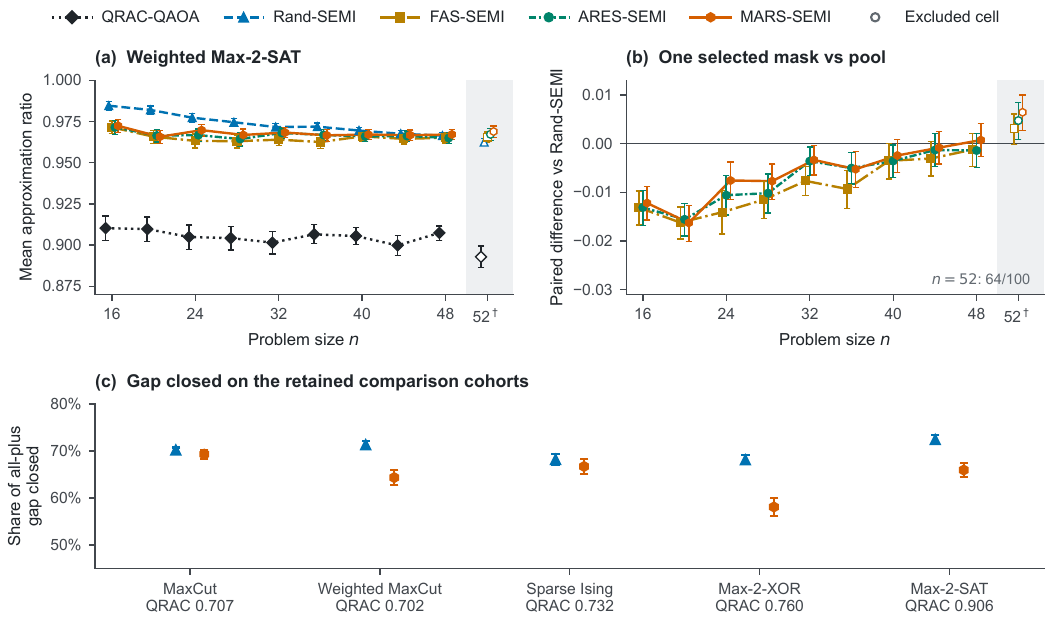}
\captionsetup{hypcap=false}
\captionof{figure}{\textbf{Weighted Max-2-SAT and the share of the all-plus gap closed.}
Noiseless $p=1$ simulations with 100 instances at each of $n=16,20,\ldots,48$.
The Max-2-SAT sizes $n\ge28$ are supplementary scaling diagnostics after
the $n=28$ surrogate-accuracy check failed, as detailed in this appendix.
Open symbols and $\dagger$ mark the censored $n=52$ cell (64 instances), which is excluded from every pooled statistic.
(a)~Mean approximation ratio.
(b)~Paired per-size differences of FAS-, ARES- and MARS-SEMI minus Rand-SEMI; positive values favor the selected mask, and the shaded region marks the incomplete cell.
(c)~Share of the QRAC-QAOA gap to the optimum that is closed, $\overline{(R_{\rm method}-R_{\rm QRAC})}/\overline{(1-R_{\rm QRAC})}$, on the retained comparison cohorts of the five unconstrained families; each label gives the all-plus mean, and the Ising score is range-normalized. Bars are pointwise 95\% instance-bootstrap intervals; in (c) the ratio of means is recomputed on each resample. Rand-SEMI optimizes $r^2$ masks per instance; every other method optimizes one.}
\label{fig:max2sat}
\end{minipage}

\end{document}